\documentclass[acmsmall,screen,review=false]{acmart}

\AtBeginDocument{%
  }

\usepackage{wasysym}
\usepackage{color,soul}
\usepackage{xcolor}
\usepackage{multirow}
\usepackage{changepage}
\usepackage{threeparttable}
\usepackage{booktabs}
\usepackage[many]{tcolorbox} 
\usepackage{fontawesome5} 
\usepackage{colortbl}
\newtcolorbox{boxK}{
    top=2pt,
    bottom=2pt,
    left=2pt,
    right=2pt,
    boxrule = 0pt,
    toprule = 0pt, 
}
\soulregister{\cite}7 
\soulregister{\citep}7 
\soulregister{\citet}7 
\soulregister{\ref}7 
\soulregister{\pageref}7 

\setcopyright{acmlicensed}
\acmDOI{10.1145/3728940}
\acmYear{2025}
\acmJournal{PACMSE}
\acmVolume{2}
\acmNumber{ISSTA}
\acmArticle{ISSTA063}
\acmMonth{7}
\received{2024-10-31}
\received[accepted]{2025-03-31}

\begin{document}


\title{ClassEval-T: Evaluating Large Language Models in Class-Level Code Translation}

\author{Pengyu Xue}
\email{xuepengyu@mail.sdu.edu.cn}
\orcid{0009-0007-3395-9575}
\authornote{These authors contributed equally to this work.}
\affiliation{%
   \institution{Shandong University}
   \city{Qingdao}
   \state{Shandong}
   \country{China}
}

\author{Linhao Wu}
\email{wulinhao@mail.sdu.edu.cn}
\orcid{0009-0001-7624-156X}
\authornotemark[1]
\affiliation{%
   \institution{Shandong University}
   \city{Qingdao}
   \state{Shandong}
   \country{China}
}
\author{Zhen Yang}
\email{zhenyang@sdu.edu.cn}
\orcid{0000-0003-0670-4538}
\affiliation{%
   \institution{Shandong University}
   \city{Qingdao}
   \state{Shandong}
   \country{China}
}
\authornote{Corresponding author.}
\author{Chengyi Wang}
\email{202300130150@mail.sdu.edu.cn}
\orcid{0009-0000-4153-6774}
\affiliation{%
   \institution{Shandong University}
   \city{Qingdao}
   \state{Shandong}
   \country{China}
}

\author{Xiang Li}
\email{leexiang@mail.sdu.edu.cn}
\orcid{0009-0007-3473-8221}
\affiliation{%
   \institution{Shandong University}
   \city{Qingdao}
   \state{Shandong}
   \country{China}
}

\author{Yuxiang Zhang}
\email{zhangyuxiang1412@mail.sdu.edu.cn}
\orcid{0009-0005-5366-9140}
\affiliation{%
   \institution{Shandong University}
   \city{Qingdao}
   \state{Shandong}
   \country{China}
}
\author{Jia Li}
\email{lijia@stu.pku.edu.cn}
\orcid{0000-0002-5579-8852}
\affiliation{%
  \institution{Tsinghua University}
  \city{Beijing}
  \country{China}
}
\author{Ruikai Jin}
\email{jrk@mail.sdu.edu.cn}
\orcid{0009-0006-9253-7906}
\affiliation{%
   \institution{Shandong University}
   \city{Qingdao}
   \state{Shandong}
   \country{China}
}

\author{Yifei Pei}
\email{peiyifei@mail.sdu.edu.cn}
\orcid{0009-0005-1351-0565}
\affiliation{%
   \institution{Shandong University}
   \city{Qingdao}
   \state{Shandong}
   \country{China}
}

\author{Zhaoyan Shen}
\email{shenzhaoyan@sdu.edu.cn}
\orcid{0000-0001-9526-6634}
\affiliation{%
   \institution{Shandong University}
   \city{Qingdao}
   \state{Shandong}
   \country{China}
}

\author{Xiran Lyu}
\email{202400130069@mail.sdu.edu.cn}
\orcid{0009-0002-8861-8907}
\affiliation{%
   \institution{Shandong University}
   \city{Qingdao}
   \state{Shandong}
   \country{China}
}
\author{Jacky Wai Keung}
\email{jacky.keung@cityu.edu.hk}
\orcid{0000-0002-3803-9600}
\affiliation{%
   \institution{City University of Hong Kong}
   \state{Hong Kong}
   \country{China}
}
\renewcommand{\shortauthors}{Xue et al.}

\begin{abstract}
In recent years, Large Language Models (LLMs) have dramatically advanced the performance of automated code translation, making their computational accuracy score reach up to over 80\% on many previous benchmarks. However, most code samples in these benchmarks are short, standalone, statement/method-level, and algorithmic, which is not aligned with practical coding tasks. Therefore, it is still unknown the actual capability of LLMs in translating code samples written for daily development.

To achieve this, we construct a class-level code translation benchmark, ClassEval-T, and make the first attempt to extensively assess recent LLMs' performance on class-level code translation. ClassEval-T is extended from ClassEval, a well-known class-level Python code generation benchmark consisting of multiple practical coding topics, such as database operation and game design, and diverse contextual dependencies (e.g., fields, methods, and libraries). It cost us 360 person-hours to accomplish the manual migration to Java and C++ with complete code samples and associated test suites. Subsequently, we design three translation strategies (i.e., holistic, min-dependency, and standalone) for class-level code translations and evaluate {eight} recent LLMs of {commercial, general, and code kinds} in diverse families and sizes on ClassEval-T. Experimental results demonstrate a remarkable performance drop compared with the most widely studied method-level code translation benchmark, and obvious discrepancies among LLMs appear, showing the effectiveness of ClassEval-T in measuring recent LLMs. Afterwards, we further discuss the usage scenarios for diverse translation strategies and LLMs' ability to dependency awareness when translating class samples. Finally, {1,243} failure cases made by the best-performing LLM under test are thoroughly analyzed and categorized in this paper for practical guidance and future enlightenment.    
\end{abstract}

\ccsdesc[500]{Software and its engineering~Empirical software validation}
\keywords{Class-Level Code Translation, Large Language Models, Benchmark}

\maketitle

\section{Introduction}
{Automated code translation seeks to efficiently migrate codebases between Programming Languages (PLs) to meet the diverse needs of various platforms, such as desktop applications, websites, and mobile apps. This is essential for enhancing coding productivity and facilitating the extension of business platforms \cite{10172589, nguyen2015divide, wu2010aura}. As software development progresses, the need to port code from one PL to another has become increasingly important.} In recent years, with the advancement of Large Language Models (LLMs), the correctness and readability of code translation have achieved substantial improvement \cite{yang2024exploring, yuan2024transagent,nitin2024spectra}. According to a series of recent studies \cite{yang2024exploring, pan2024lost}, LLMs, such as CodeLlama \cite{roziere2023code} and GPT-3.5 \cite{chatgpt}, can correctly translate over {70\%-80\% of code samples} with the most basic prompt among diverse translation pairs, even for those domain-specific scenarios \cite{tao2024unraveling}, such as Python-to-Scala. 
{However, previous studies in code translation primarily experiment with statement/method-level code translation benchmarks, which typically involve shorter code snippets, limited dependencies, and a focus on algorithmic problems. These benchmarks, while useful, fail to capture the complexities encountered in real-world software development. In practice, code translation tasks often involve longer code structures, more intricate dependencies, and diverse functionalities. Therefore, the current benchmarks do not adequately reflect the code translation capabilities of LLMs in real-world coding scenarios. This gap highlights the urgent need for a brand-new code translation benchmark aligning with coding practice for LLMs' assessment, thereby providing insightful research direction and practical guidance for academia and industry.}

\textbf{Benchmark ClassEval-T:} To mitigate the above limitation in code translation assessment, we construct ClassEval-T, a class-level code translation benchmark comprising three parallel PLs (i.e., Python, C++, and Java) and associated test suites with extremely high coverage scores of 99.7\% on statements and 98.2\% on branches. ClassEval-T is extended from the latest class-level Python code generation benchmark, ClassEval \cite{du2024evaluating}, via line-wise manual translation for 360 person-hours. Hence, ClassEval-T also contains the virtues that ClassEval has, including longer code lengths, diverse dependencies on fields, methods, and libraries, as well as practical coding problems (e.g., database operation and game design). Besides, a class-level code translation benchmark stands out from those statement/method-level ones in two ways. (1) ClassEval-T not only can evaluate the correctness of translated code but also can assess LLMs' ability in dependencies awareness and inference on proper library invocation. (2) Class-level code translation allows for further exploration of diverse translation strategies (e.g., translating the whole class at once or separately).

\textbf{Empirical Study:} Based on ClassEval-T, we make the first attempt to extensively assess recent LLMs’ performance on class-level code translation. {Specifically, eight recent LLMs of diverse kinds (e.g., commercial, general, and code) and sizes (ranging from 7B to 671B) are involved.} For each studied LLM, we evaluate their performance with three distinctive translation strategies, i.e., holistic translation (translating the whole class at once), min-dependency translation (translating each module of a class one by one, given the minimal necessary dependencies), and standalone translation (translating each module of a class one by one without dependency information). For each translated code sample, we evaluate its correctness via Computational Accuracy (CA) and Compilation Success Rate (CSR). Besides, we also incorporate DEPendency evaluation metrics (DEP) to investigate LLMs' awareness of necessary context during the translation for class-level samples and thoroughly analyze and categorize {1,243} failed cases at the end of this experiment.

\textbf{Main findings:} Based on our results, we have the following findings. (1) All LLMs perform dramatically worse on class-level code translations than method-level ones, where the latter cannot even tell different LLMs' performance discrepancies. (2) {Commercial LLMs (e.g, DeepSeek-V3, GPT-4o, and Claude-3.5-Sonnet) exhibit predominately superior performance on class-level code translation, while smaller LLMs normally perform worse, given other factors the same.} Besides, All LLMs perform better on Python-oriented translations, and code LLMs do not necessarily outperform general LLMs. (3) {Commercial LLMs always work better on the holistic translation strategy, while the selection of translation strategy for smaller LLMs depends on different scenarios. Specifically, as for Python- and Java-oriented translations, smaller LLMs work better on the holistic strategy, while for C++-oriented translations, they perform neck-to-neck on holistic and min-dependency strategies.} (4) The holistic translation strategy improves LLMs’ awareness of field dependencies, while the min-dependency strategy allows LLMs to invoke more necessary libraries. As for method dependencies, the above two strategies perform neck-to-neck. (5) Syntax errors remain a primary issue in class-level translations, but class-related errors, such as function/variable usage and consistency issues, are more prominent in C++/Java-oriented translations. Based on the above findings, several implications are summarized for both researchers and practitioners. The contributions of this work can be three-fold:

(1) We manually construct the first class-level code translation benchmark. The benchmark dataset and associated code are available at \cite{wLinHooC78:online}.

(2) We conduct the first attempt to extensively evaluate diverse LLMs' class-level code translation capability with different translation strategies and assessing aspects. 

(3) We summarize a series of findings and manually analyze {1,243} failed cases for categorization, shedding light on practical guidance and future research directions.

\section{Related Work}

\subsection{Large Language Models for Code Translation}

Large Language Models (LLMs), with more than billions of parameters trained on general textual/code corpora and instructions, have achieved state-of-the-art performance in many coding tasks \cite{xue2024exploring,ahmed2022few, xue2024automated,DevEval,AceCoder,SCoT,aiXcoder-7B,EvoCodeBench}. In recent years, owing to the great demand for codebase migrations \cite{yang2024vert, tao2024unraveling}, automated code translation with LLMs has also attracted increasingly substantial attention. 
For example, Pan et al. \cite{pan2024lost} investigated the performance of five LLMs in code translation tasks and identified and summarized syntax and semantic errors in their translations. Eniser et al. \cite{eniser2024towards} presented a comprehensive study on LLM-based translation to Rust, where five LLMs are evaluated.
Additionally, other studies have explored ways to enhance LLM performance in code translation tasks. For example, Yang et al. \cite{yang2024exploring} introduced UniTrans, which utilizes test cases to improve LLM translation performance. Yuan et al. \cite{yuan2024transagent} proposed a novel LLM-based multi-agent system called TRANSAGENT, which enhances LLM-based code translation by correcting syntax and semantic errors through the synergy of four LLM-based agents. {Pan et al. {\cite{pan2023stelocoder}} introduced SteloCoder, a decoder-only StarCoder-based {\cite{li2023starcoder}} LLM designed specifically for Python-oriented translation.} Yin et al. \cite{yin2024rectifier} proposed a general corrector, Rectifier, a lightweight and universal model for repairing translation errors by learning from mistakes made by existing LLMs. However, despite these studies demonstrating the promise of using LLMs for code translation, their evaluations have primarily been conducted on statement- or method-level benchmarks, 
leaving a notable lack of studies on code translations for more challenging scenarios.
Considering most practical coding is object-oriented \cite{du2024evaluating}, this work constructs a class-level benchmark, ClassEval-T, with longer code lengths, diverse dependencies, and practical coding problems for investigation.

\subsection{Existing Benchmarks for Code Translation}

Code translation plays a critical role in cross-language codebase migration. In recent years, a series of benchmarks were successively constructed,
typically providing parallel corpora of diverse PLs with accompanied test suites for correctness validation. For example, Puri et al. \cite{puricodenet} built CodeNet, which covers widely used PLs and includes code translation corpus collected from programming contest sites with extensive metadata. AVATAR \cite{ahmad2023avatar} is a collection of programming problems with solutions written in Java and Python, sourced from competitive programming sites, online platforms, and open-source repositories. TransCoder-test \cite{roziere2020unsupervised} is the most extensively assessed code translation test set consisting of 948 parallel samples in C++, Java, and Python, derived from GeeksForGeeks\cite{geeksforgeeks}. However, it only contains 568 samples with test suites in at least one PL. Owing to its huge noise, Yang et al. \cite{yang2024exploring} cleaned this dataset and released a new version. Building on existing datasets, Jiao et al. \cite{jiao2023evaluation} constructed G-TransEval by extracting parallel functions from program-level parallel code, covering five PLs across four different types. 
{Yan et al. \cite{yan2023codetransocean} constructed CodeTransOcean, a large-scale comprehensive benchmark, where the MultilingualTrans dataset supports translation between eight popular PLs. However, even though the MultilingualTrans test set contains over 7,000 samples, after our filtering process, only 44 tasks simultaneously cover all parallel PLs and primarily focus on the translation of basic syntactic structures to algorithmic implementations. xCodeEval \cite{khan2024xcodeeval} is an execution-based multilingual multitask evaluation benchmarks, derived from data collected from Codeforces \cite{Codeforc25:online}, and can be utilized for tasks such as code generation and code translation.} Besides, Tao et al. \cite{tao2024unraveling}, focusing on more diverse and domain-specific PLs, constructed PolyHumanEval, which extends HumanEval into a multilingual benchmark across 14 PLs. Table \ref{benchmark} provides an overview of the above benchmarks, including their release time, parallel PLs, whether they align with practical development, granularity, 
benchmark size (\#Tasks), {average number of methods per PL (\#Methods/P)}, the average number of test cases per task (\#Tests/T), code scale (\#LOC/T: average lines of code per task, \#Tokens/T: average number of tokens per task, split by blanks), and the number(ratio) of diverse dependencies (\#FD: Field dependencies, \#MD: Method dependencies, \#LD: Library dependencies) across different benchmarks. 
We also present our constructed benchmark ClassEval-T in the last row for comparison.

{Additionally, there are some benchmarks for code translation tasks that do not include parallel PLs. For example, Pan et al. \cite{pan2024lost} manually constructed equivalent Java test suites via \textit{JUnit}\cite{junit5} from the Python test suites implemented in EvalPlus \cite{liu2024your}, a code synthesis evaluation framework extended from HumanEval \cite{chen2021evaluating}, for the validation of Python-to-Java translation. The NicheTrans test set in CodeTransOcean \cite{yan2023codetransocean} focuses on translations from domain-specific PLs to popular PLs. Yan et al. \cite{yan2023codescope} introduced CodeScope, an execution-based, multilingual, multitask evaluation benchmark crawled from Codeforces for measuring LLM capabilities on coding tasks including code translation. Furthermore, there are also two concurrent works focusing on repository-level code translation. Ibrahimzada et al. \cite{ibrahimzada2024repository} proposed AlphaTrans to translate ten real-world open-source projects from Java to Python. Ou et al. \cite{ou2024repository} introduced a repository-level code translation benchmark comprising 375 tasks targeting Rust. However, owing to the absence of parallel PLs in these benchmarks, two main limitations appear. Firstly, researchers cannot fairly compare models' translation performance among different PLs, as samples of different PLs implementing different functionalities. Secondly, parallel samples can be considered as an assurance that samples can be translated from one PL to another. Without parallel samples, untranslatable samples may present in the test set. Once models fail in certain cases, we cannot identify whether it is owing to models' weaknesses or the untranslatability of cases, thereby inducing biased and unreliable evaluations.
As a result, we consider the above benchmarks to be flawed for evaluation, and we only include and report benchmarks that contain parallel PLs for comparison in Table \ref{benchmark}.}

Based on Table \ref{benchmark}, we have the following observations.
\textbf{ 
First, existing benchmarks typically focus on statement- or method-level code translation tasks (Column ``Granularity'').} {Thus, such tasks often involve a limited number of lines (21.8 on average) and tokens (95.2 on average), which may not fully explore the capacity of recent LLMs that can process longer sequences.} 
\textbf{Second, statement- or method-level code units lack field and method dependencies (Column ``\#FD'' and ``\#MD'') or provide minimal library dependencies (Column ``\#LD'')}. Because they are normally crawled from online judge websites, such as AtCoder \cite{AtCoder51:online}, Codeforces \cite{Codeforc25:online}, and GeeksForGeeks \cite{geeksforgeeks}, leading them to mainly focus on translating standalone code units without consideration of other code contexts. However, in practical scenarios, methods often depend on each other or share variables. Previous study \cite{yu2024codereval} indicated that in open-source projects, only about 30\% of methods operate independently of other code contexts. Therefore, the capability of LLMs to translate class-level code involving interdependent methods remains unclear.

To address this gap, we manually construct the first class-level code translation benchmark, ClassEval-T. Compared to existing benchmarks, ClassEval-T includes more lines, tokens, and test suites. Furthermore, it provides richer dependency information, establishing ClassEval-T as the first comprehensive benchmark to evaluate LLMs' abilities in translating longer, interdependent, class-level code snippets.

\begin{table}[]
\vspace{-1em}
\setlength{\abovecaptionskip}{0cm}
\renewcommand{\arraystretch}{1.4}
\caption{Existing Benchmarks for Code Translation}
\label{benchmark}
\scriptsize
\setlength\tabcolsep{1.7pt}
\begin{threeparttable}
\resizebox{\textwidth}{!}{ 
\begin{tabular}{cccccccccccccc}
\toprule
Benchmark                                               & Time          & Parallel Programming Languages                                                                                                            & Practical  & Granularity            & \#Tasks     & \#Methods/P    & \#Tests/T     & \#LOC/T       & \#Tokens/T     & \#FD                 & \#MD                 & \#LD                \\ \hline
CodeNet \cite{puricodenet}             & 2021          & C,C++,Go,Java,Python                                                                                                                      & N          & Statement/Method-level & 200         & 111          & 1             & 34.9          & 112.0          & -                    & -                    & 21(18.9\%)          \\ \hline
AVATAR \cite{ahmad2023avatar}          & 2021          & Java,Python                                                                                                                               & N          & Statement/Method-level & 250         & 224          & 25.1          & 26.9          & 149.5          & -                    & -                    & 51(22.8\%)          \\ \hline
G-TransEval \cite{jiao2023evaluation}  & 2023          & C++,C\#,Java,Python,JavaScript                                                                                                             & N          & Method-level           & 400         & 400          & 5             & 14.3          & 93.6           & -                    & -                    & -                   \\ \hline
\begin{tabular}[c]{@{}c@{}}CodeTransOcean-\\ MultilingualTrans \cite{yan2023codetransocean}\end{tabular}                                     & 2023          & \begin{tabular}[c]{@{}c@{}}C,C++,C\#,Java,Python,\\ Go,PHP,Visual Basic\end{tabular}                                                       & N          & Statement/Method-level & 44          & 57.9             & 1             & 23.8          & 82.1           & -                    & 8(27.6\%)                    & 3(10.3\%)                   \\ \hline
xCodeEval-test \cite{khan2024xcodeeval}                                         & 2024          & \begin{tabular}[c]{@{}c@{}}C,C++,C\#,Java,Python,Ruby,Go,\\ JavaScript,Kotlin,PHP,Rust\end{tabular}                                        & N          & Statement/Method-level & 226         & 406.1          & 46.8          & 30.8          & 95.5           & -                    & 24(20.2\%)                    & 27(20.1\%)          \\ \hline
Yang et al. \cite{yang2024exploring}   & 2024          & Python,Java,C++                                                                                                                           & N          & Method-level           & 568         & 568          & 6.2           & 12.4          & 95.8           & -                    & -                    & -                   \\ \hline
PolyHumanEval \cite{tao2024unraveling} & 2024          & \begin{tabular}[c]{@{}c@{}}Python,C++,C\#,Dart,Go,Java,JavaScript,PHP,\\ Kotlin,Ruby,Rust,Scala,Swift,TypeScript\end{tabular}             & N          & Method-level           & 164         & 164          & 8.1           & 9.6           & 37.9           & -                    & -                    & 7(4.3\%)            \\ \hline
\textbf{ClassEval-T}                                    & \textbf{2024} & \textbf{Python,Java,C++}                                                                                                                   & \textbf{Y} & \textbf{Class-level}   & \textbf{94} & \textbf{386} & \textbf{33.8} & \textbf{66.7} & \textbf{199.5} & \textbf{252(65.3\%)} & \textbf{100(25.9\%)} & \textbf{74(19.2\%)} \\ \bottomrule
\end{tabular}}
\vspace{-1.5em}
\end{threeparttable}
\end{table}

\section{New Benchmark: ClassEval-T}

This section introduces the new benchmark, namely ClassEval-T, which contains three parallel PL corpus (i.e., Python, Java, and C++) of class-level code snippets. {We illustrate the construction} procedure and superiority of ClassEval-T below.
\subsection{Benchmark Construction}
\subsubsection{Calibration Selection}
Constructing a brand-new code translation benchmark from scratch is extremely time-consuming and complicated because it is hard to crawl semantic-equivalent code snippets of different PLs in the wild \cite{roziere2020unsupervised, roziere2021leveraging}, not to mention the line-wise consistency and class-level code snippets required in this work. 
Therefore, this work adopts a strategy that leverages a code generation benchmark as calibration and then manually translates those samples from one PL to others, which can also avoid the notorious data leakage risk during the evaluation. After revisiting a series of existing code generation benchmarks, we select ClassEval \cite{du2024evaluating} as our calibration dataset, because (1) it is the latest manually crafted code generation benchmark exclusively for Python code on the class level, (2) it covers a wide range of topics in practical software development (e.g., management system and database operations), (3) it constructs sufficient test suites from both method-level and class-level with high test coverage (e.g., 98.2\% and 99.7\% of branch-level and statement-level coverage), and (4) it involves multiple instance methods for each class and diverse dependencies to fields, methods, and third-party libraries. As shown in Table \ref{benchmark}, all of the current code translation benchmarks ignore almost all the above attributes, leading to the inability to effectively assess and unveil the performance discrepancies among LLMs. We also noticed some other repo-level code generation benchmarks, such as CoderEval \cite{yu2024codereval} and DevEval \cite{li2024deveval}, consisting of class-level characteristics as well. Nevertheless, we do not select them as the calibrations because most of their samples involve interdependence among multiple files and classes, making it hard to discriminate enough standalone classes for evaluation. In the following subsections, we define principles for manually translating Python code snippets in ClassEval to their corresponding Java and C++ versions.
\subsubsection{Construction Principles}
To ensure manually translated code snippets are of high quality, we require annotators to rigorously translate Python programs and test suites of ClassEval to corresponding counterparts of Java and C++ line by line. Furthermore, to ensure the consistency of translation, we also define five translation principles, including naming conventions, type conversion, implementation layout, library selection, and test suite construction.

\textbf{Principle 1 (naming convention):} According to literature and coding practice \cite{reddy2000java, li2022ropgen, sutter2004c++}, we define a series of naming convention rules for various elements in Java and C++ programs. For example, classes are required to be named in the pascal case (e.g., \textit{AreaCalculator}) while constants are required to be named in the screaming snake case (e.g., \textit{M\_PI}) for both Java and C++. Packages only exist in Java programs with only lower-case letters (e.g., \textit{package org.example;}), while Structs are special data structures in C++ with pascal case. For naming conventions of fields, methods, and variables in Java are imposed on camel case (e.g., \textit{totalNum}), while those in C++ are cast in snake case (e.g., \textit{check\_time}).

\textbf{Principle 2 (type conversion):} Python, as a dynamic-typed PL, does not explicitly declare variable/constant types in source code but relies on type inference at runtime, which is handled by its interpreter. Thus, we need to determine a uniform standard to convert variable/constant types in Java and C++ sides, as they are static-typed PLs. Specifically, we consistently use type \textit{double} to replace real numbers used in Python programs while using type \textit{int} to replace integers. Besides, for Python variables with dynamic assignments, we substitute them with type \textit{auto} and \textit{Object} in C++ and Java, respectively. The former makes C++ compilers deduce types for variables automatically while \textit{Object} in the latter scenario can be assigned by variables of any data type.  

\textbf{Principle 3 (implementation layout):} In addition to the above principles, we also make specific criteria for implementation layout during the manual translation. Both translated Java and C++ programs are required to use K\&R brace style and 4 spaces per indentation level \cite{reddy2000java, sutter2004c++}. Besides, classes associated with function declarations are listed in the same files with function implementations for C++. We also removed all original comments in ClassEval to ensure there are no extra hints for follow-up translation. 

\textbf{Principle 4 (library selection):}
It is very common to encounter programs with library dependencies during the manual translation. For build-in libraries of Python, such as \textit{math}\cite{pydoc} and \text{os}\textsuperscript{\cite{pydoc}}, there are explicit one-to-one mapping to Java (i.e., \textit{java.lang.Math}\cite{javadoc} and \textit{java.lang.System}\textsuperscript{\cite{javadoc}}) and C++ (i.e., \textit{cmath}\cite{cppdoc} and \textit{cstdlib}\cite{cppdoc}). However, for Python code involving Third-Party Libraries (TPLs), one-to-one mapping TPLs may not exist to implement complete functionalities in Java and C++. From a conciseness and efficient perspective, we require annotators to adequately search the specific TPLs on programming Q\&A websites (e.g., Stack Overflow\cite{stackoverflow} and Quora\cite{quora}) and ChatGPT\cite{chatgpt}, thereby obtaining minimized TPL sets to accomplish the functionality equivalency. To better manage the associated libraries, \textit{maven}\cite{maven} and \textit{nuget}\cite{nuget} are required to be used for Java and C++, respectively, in this work. 

\textbf{Principle 5 (test suite construction):} ClassEval adopted \textit{unittest}\cite{unittest} to construct both method-level and class-level test suites. It is a unit testing framework exclusively for Python programs with various assertion APIs and Test Fixtures (e.g., setUp and tearDown methods) to prepare and clean up tasks before and after test execution. Accordingly, we select \textit{JUnit}\cite{junit5} and \textit{GoogleTest}\cite{googletest}, two widely used testing frameworks designed for Java and C++ with similar functionalities, respectively, to completely translate those test suites to Java and C++ sides.

\subsubsection{Construction Procedure:} Following the aforementioned principles, five authors with over three years of coding experience in Python, Java, and C++ are engaged in the manual translation procedure. Among them, one individual serves as the lead, responsible for review and arbitration, while the remaining four authors are uniformly divided into two groups. Each pair translates Python programs and associated test suites of ClassEval to Java or C++ sides separately, where one annotator is responsible for writing translated programs and test suites, and the other one for double-checking and executing the translated results, ensuring translation quality and correctness. Once they encounter disagreement on translation design or logic, the lead participates in the discussion and facilitates them to reach a consensus, adhering to the above principles. The whole benchmark construction consumes 360 person-hours. Finally, except for six samples containing third-party libraries without C++/Java versions, such as NLTK \cite{NLTKNatu1:online} and PyPDF2 \cite{Welcomet83:online}, 94 parallel class-level coding tasks are constructed across main PLs, including Python, Java, and C++.
\subsection{The superiority of ClassEval-T} 
ClassEval-T features a lot of superiority over its previous counterparts, where its detailed statistics are shown in the last row of Table \ref{benchmark}. (1) It consists of 94 different practical coding tasks manually designed by Du et al. \cite{du2024evaluating}, including the student registration system, area calculator, gomoku game, etc. (2) {As an evaluation benchmark for code translation, ClassEval-T contains more lines (66.7) and tokens (199.5) per task, outperforming previous benchmarks on average by 205.96\% and 109.56\% respectively.} (3) Besides, ClassEval-T carries more diverse dependencies to fields, methods, and libraries. As shown in Table \ref{benchmark}, 65.3\% of the code samples in ClassEval-T contain field dependencies, 25.9\% of them contain method dependencies, and 19.2\% of them contain library dependencies. (4) {To ensure the effectiveness of validation, ClassEval-T comprises 33.8 test cases per task, outperforming previous benchmarks by 154.14\%.} At the same time, it also carries extremely high coverage scores on statements (99.7\%) and branches (98.2\%). 
Consequently, ClassEval-T demonstrates a much higher complexity and comprehensive test suites that can be used for the assessment of state-of-the-art LLMs.

\section{Experimental Design}
This study uses ClassEval-T to evaluate recent LLMs on class-level code translation with the following Research Questions (RQs).

\textbf{RQ1 (Overall Correctness): How do recent LLMs perform on class-level code translation?} Considering longer lines of code, diverse dependencies, and more knowledge-intensive, especially on TPL selection, class-level code translation carries much more value to be investigated compared with method-level code translation.

\textbf{RQ2 (Translation Strategies): How do different translation strategies affect the performance of recent LLMs?} In practice, developers may use different translation strategies according to their specific requirements. Therefore, exploring different translation strategies is vital for practical guidance.

\textbf{RQ3 (Dependency Awareness): To what extent can LLMs translate code dependent on other contexts in class-level code translation?} Class-level code translation involves frequent and diverse dependencies on fields, methods, and libraries, while properly importing and leveraging contextual dependencies during class-level code translation is critical to make the translated programs of high readability and maintainability. Hence, we investigate the above ability of LLMs in this RQ.

\textbf{RQ4 (Failed Cases Analysis): What kind of errors will LLMs make and how frequent they are?} The weaknesses of LLMs in method-level code translation have been fully investigated in recent years, and a series of workarounds were proposed to remedy them \cite{yang2024exploring,tao2024unraveling,pan2024lost, yuan2024transagent}. Nonetheless, it is still unknown what kinds of errors will LLMs make, given more challenges in class-level code translation. Consequently, it is imperative to dig deeper into this field and try to enlighten future research.

\subsection{Studied LLMs}
{
Table \ref{Studied LLMs in this paper} lists our studied LLMs with their categories, model type, released times (Time), parameter sizes (Size), pretraining token numbers (Training Base), and the input/output context window token limits (In/Out). For those models whose context windows are measured in characters, we specified them separately.} {As can be seen, to extensively explore the performance of latest LLMs, we select LLMs from general (i.e., Llama 3 \cite{Llama3268:online} and Gemma \cite{team2024gemma}), code (i.e., CodeLlama \cite{roziere2023code} and CodeGemma \cite{team2024codegemma}), and commercial kinds \footnote{We define commercial LLMs as large language models with substantial parameter sizes  (typically exceeding 100B), developed by organizations for commercial purposes, and providing publicly accessible interfaces for interactive dialogue and text-based applications.} (i.e., DeepSeek-V3 \cite{DeepSeek19:online}, GPT-4o \cite{OpenAI59:online} and Claude-3.5-sonnet \cite{Claude3555:online}), released from late 2023 to late 2024.} Furthermore, for LLMs from the same family, we also select two different sizes (i.e., Llama3-8B and Llama3-70B) to study the influence of the volume of model parameters.

\begin{table}[]
\setlength\tabcolsep{4pt} 
\vspace{-1em}
\setlength{\abovecaptionskip}{0cm}
\caption{Studied LLMs in this paper}
\label{Studied LLMs in this paper}
\scriptsize
\begin{threeparttable}
\begin{tabular}{lcccccc}
\toprule
Category                     & Model                                       & Model Type             & Time                     & Size & Training Base (Tokens) & In/Out (Tokens) \\ \midrule
\multirow{3}{*}{Commercial}                   & DeepSeek-V3 \cite{DeepSeek19:online}                               & Instruction Model & 2024-12                  & 671B & 14.8 trillion & 64k / 4096 \\ 
     & Claude-3.5-Sonnet \cite{Claude3555:online}         & Instruction Model & 2024-06 & $\sim$175B \cite{abacha2024medec} & / & 128k / 4096 \\
                             & GPT-4o \cite{OpenAI59:online}                             & Instruction Model & 2024-05 & $\sim$200B \cite{abacha2024medec} & / & 128k / 4096 \\

\midrule
\multirow{3}{*}{General} & Llama3 \cite{Llama3268:online} & Instruction Model   & 2024-04 & 8B   & 15.6 trillion & 7k / 1024 \\
                             & Llama3 \cite{Llama3268:online} & Instruction Model   & 2024-04 & 70B  & 15.6 trillion & 7k / 1024 \\
                             & Gemma \cite{team2024gemma}                      & Base Model & 2024-02 & 7B   & 6 trillion    & 11200 Characters/ 1024 \\ \midrule
\multirow{2}{*}{Code}    & CodeLlama  \cite{roziere2023code}                                 & Instruction Model  & 2023-08                  & 7B   & 500 billion   & 8000 Characters/ 1024 \\
                             & CodeGemma \cite{team2024codegemma}                                   & Base Model & 2024-04                  & 7B   & 500 billion  & 11200 Characters/ 1024 \\ 
\bottomrule
\end{tabular}
\begin{tablenotes}
\definecolor{light cyan}{HTML}{E1FFFF}
\fboxsep1.5pt
\item[$\Phi$] \scriptsize {In the following sections, ``DeepSeek'' refers to DeepSeek-V3 and ``Claude'' indicates Claude-3.5-Sonnet. We also use ``Llama3-70B'' and ``Llama3-8B'' to denote the two versions of Llama3.}
\end{tablenotes}
\end{threeparttable}
\vspace{-2em}
\end{table}

\subsection{Studied Translation Strategies}
\label{Studied Translation Strategies}
Given a class-level code translation task, we investigate the following three translation strategies for each of the studied LLMs.

\textbf{(1) Holistic Translation:} LLMs are required to translate the whole class at once, given the class-level code snippet of the source PL as inputs. This strategy aims to simulate those class-level code translations entirely from scratch.

\textbf{(2) Min-Dependency Translation:} LLMs are dictated to translate class skeletons\footnote{We define class skeletons as classes without any instance/static methods but with library dependencies, constructors, and fields.} and their corresponding methods one by one, given these items with their minimal necessary dependencies (e.g., class skeletons, fields, methods, and libraries) of the source PL as inputs sequentially. After a series of translations above, all translated results are composed together to form a complete class of the target PL. This strategy aims to mimic scenarios that only need partial translations for a class module. 

\textbf{(3) Standalone Translation:} This strategy is the same as the min-dependency translation, except we do not provide dependencies for reference but only a standalone class skeleton or method of source PL, thereby exploring the effects of dependencies in class-level code translations.

We present an example for each of the above translation strategies in our released repository for a more explicit illustration \cite{wLinHooC78:online}. Considering that code translation is a task that translates code from source PLs to target PLs, no matter which strategy we use. Thus, we design a most basic prompt to fit the above three strategies and mimic the easiest and most prevalent usage in practice, which can be formally defined as: ``Translate the following \$\{$pl_{src}$\} code to \$\{$pl_{dst}$\}:\textbackslash n\textbackslash n\$\{$prog_{src}$\}'', where \$\{$pl_{src}$\}, \$\{$pl_{dst}$\}, and \$\{$prog_{src}$\} are placeholders for the source PL (e.g., Python, Java, and C++), target PL, and a program written in a certain source PL. 

\subsection{Evaluation Metrics}
\label{Evaluation Metrics}
For translation correctness, we follow most previous studies \cite{roziere2021leveraging,roziere2020unsupervised,yang2024exploring} to measure two main aspects, including the Compilation Success Rate (CSR) and Computational Accuracy (CA).

\textbf{Compilation Success Rate (CSR):} It computes the ratio of samples that can be successfully compiled after translation. We define CSR as below.

\begin{equation}
    CSR = \frac{\sum^{N_{x}}_{k=1}cs(\hat{y_{k}})}{N_{x}}, \ \textbf{where} \ cs(\hat{y_{k}}) = \left\{\begin{matrix}
1 & Compile(\hat{y_{k}}) \to success\\
0 & Compile(\hat{y_{k}})  \to error\\
\end{matrix}\right. 
\end{equation}
where $N_{x}$ denotes the total number of samples, $\hat{y_{k}}$ denotes the $k$-th translated sample via a certain LLM. \textit{Compile($\cdot$)} denotes compiling samples with their corresponding compilation programs, such as \textit{javac} for Java while \textit{g++} for C++, and its results are either \textit{success} or \textit{error}.

\textbf{Computational Accuracy (CA):} It computes the ratio that the translated programs can produce the same execution result as the ground truths, given the same inputs. We define CA as below.
\begin{equation}
CA = \frac{\sum^{N_{x}}_{k=1}ca(y_{k},\hat{y_{k}})}{N_{x}}, \ \textbf{where} \ ca(y_{k},\hat{y_{k}}) = \left\{\begin{matrix}
1 & Exec_{k}(y_k)=Exec_{k}(\hat{y_{k}})\\
0 & Exec_{k}(y_k)\neq Exec_{k}(\hat{y_{k}}) \\
\end{matrix}\right. \label{eq2}
\end{equation}
where 
$y_k$ denotes the ground truth of the $k$-th
sample, 
$Exec_{k}(\cdot)$ denotes the execution result of a program with the test suite of the $k$-th sample.  Following the evaluation procedure of ClassEval \cite{du2024evaluating}, we compute CA on both class-level (CA$_c$) and method-level (CA$_m$), where CA$_c$ considers code samples at the class granularity, and CA$_m$ considers them at method granularity. A class-level sample is deemed correct if it passes the test suites of both class-level and method-level, while a method-level sample is deemed correct if it passes all method-level test suites. 

\textbf{DEPendency evaluation (DEP):} Apart from correctness evaluation,  
we follow \cite{du2024evaluating,yu2024codereval} and design $DEP_{F}$,$DEP_{M}$, $DEP_{L}$ to assess the ability of LLMs to be aware of diverse contextual dependencies (i.e., fields, methods, and libraries) when translating code. Specifically, they count the recall of unique dependencies that translated programs invoke and simultaneously appear in the program of source PL. We define $DEP_{X}$ below to illustrate each of the metrics above, where $X$ can be $F$/$M$/$L$ for fields, methods, and libraries, respectively:

\begin{equation}
    DEP_{X} = \frac{1}{N}\sum_{i=1}^{i=N}{\frac{\sum_{j=1}^{j=J}d_{X}(\hat{m_{ij}}^{dst}) \cap d_{X}(m_{ij}^{src})}{\sum_{j=1}^{j=J}d_{X}(m_{ij}^{src})}}
\end{equation}

where $N$ denotes the total number of tasks (classes), $J$ denotes the number of methods of the $i$-th class sample, $m_{ij}^{src}$ denotes the $j$-th method of the $i$-th class in source PL, while $\hat{m_{ij}}^{dst}$ denotes the translated result of method $m_{ij}^{src}$.
Besides, $d_{X}(\cdot)$ count the number of dependencies on fields/methods/libraries of the method $m_{ij}^{src}$ or $\hat{m_{ij}}^{dst}$, while $d_{X}(\hat{m_{ij}}^{dst}) \cap d_{X}(m_{ij}^{src})$ counts the number of those dependencies both appear in $m_{ij}^{src}$ and $\hat{m_{ij}}^{dst}$.
Detailed measurement workarounds for $d_{X}(\cdot)$ for fields/methods/libraries are illustrated in Section \ref{RQ3: Dependency Awareness}.
\subsection{Implementation Details}

{We utilize eight LLMs for our experiments. Specifically, we employ the Deepseek, GPT-4o and Claude via their official API interface\cite{deepseek,OpenAI59:online,Claude3555:online}.} Additionally, CodeGemma, Gemma, and Llama3-70B are accessed through the open API interface provided by NVIDIA\cite{nvidia}, while Llama3-8B and CodeLlama are operated via the open API interface provided by Baidu Qianfan platform\cite{qianfan}. The hyperparameters for the code translation process are set as follows: $temperature = 0.8$, $top\_p = 0.95$, $top\_k = 50$, and $n = 1$, which means we only fetch LLMs' first translated candidates for evaluation during inference. All other hyperparameters are kept by default. {To mitigate the impact of LLM randomness on result reliability, following prior works{\cite{liu2024your}\cite{yan2023codetransocean}}, we repeat the experiments three times under the same settings and report the averaged results. All evaluations are performed in a zero-shot setting.

}


\section{Experimental Results}
\subsection{RQ1: Overall Correctness}
\begin{figure}[htbp]
\vspace{-1em}
\setlength{\abovecaptionskip}{0cm}
\includegraphics[width=0.8\textwidth]{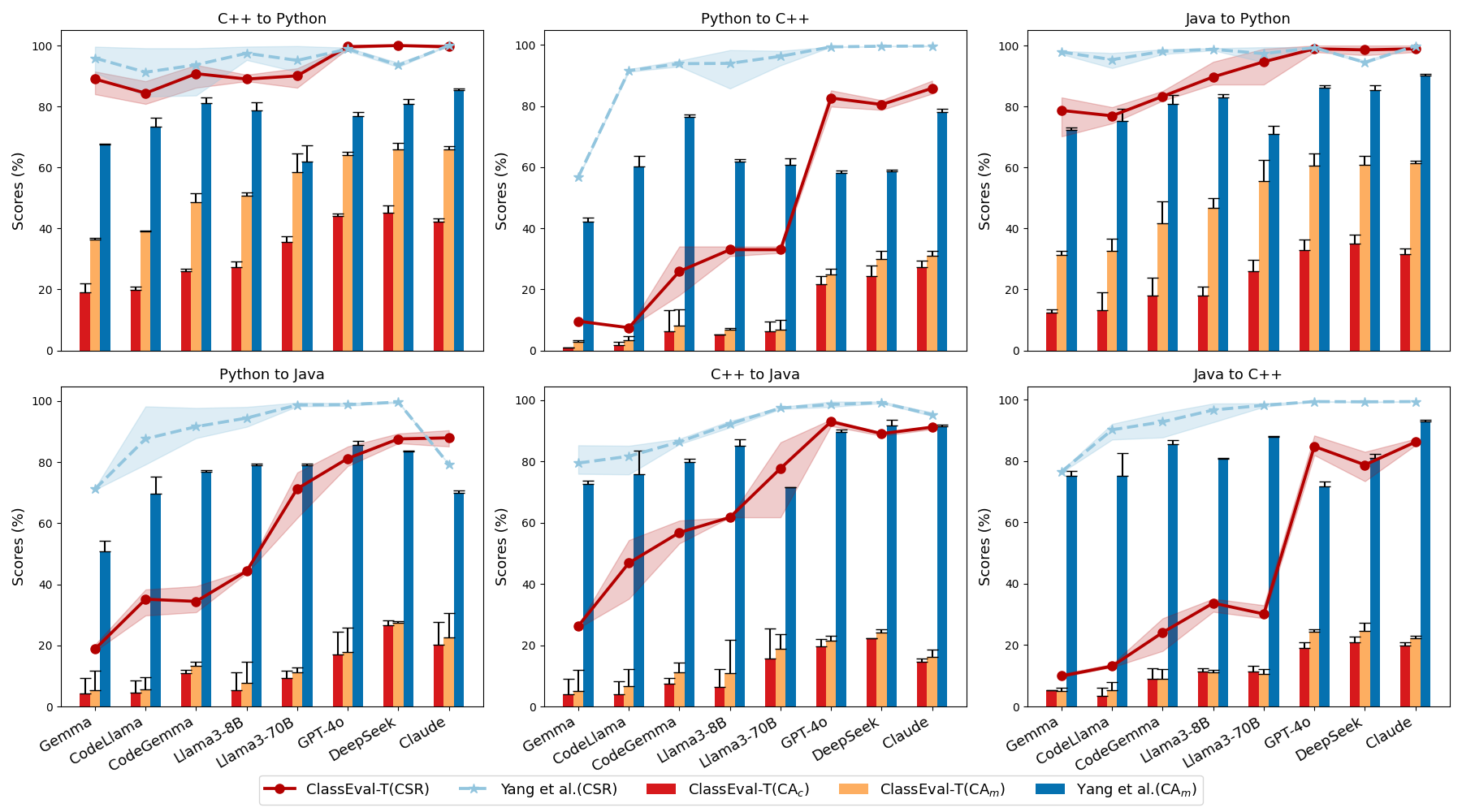}
\caption{Translation results of CSR, CA\(_c\) and CA\(_m\) on benchmarks of Yang et al. and ClassEval-T}
\label{comparison}
\vspace{-2em}
\end{figure}

Apart from the experiments on ClassEval-T, we also evaluate each studied LLM on Yang et al.'s code translation dataset \cite{yang2024exploring}, which is cleaned from \cite{roziere2020unsupervised}, one of the most widely used method-level datasets in the code translation field \cite{roziere2021leveraging, 10172589,szafraniec2023code}. Figure \ref{comparison} demonstrates the correctness comparison between code translation benchmarks of ClassEval-T and Yang et al. among diverse LLMs in terms of average CA and CSR, where CA is added with error bars of 95\% confidence interval while CSR is wrapped with ranges among repeated experiments. In particular, the holistic translation strategy is adopted here, and detailed strategy comparison results are showcased in Section \ref{RQ2: Translation Strategies}. As can be seen, both class-level CA (CA$_c$) and method-level CA (CA$_m$) are adopted for assessment on ClassEval-T, while Yang et al.'s dataset is solely evaluated on CA$_m$ as it only consists of method-level code translation samples. Table \ref{rq1} presents specific experimental results with the holistic translation strategy on ClassEval-T. Based on Table \ref{rq1} and Figure \ref{comparison}, we have the following observations.

\begin{table}[]
\vspace{-1em}
\setlength{\abovecaptionskip}{0cm}
\caption{Experimental Results with holistic translation on ClassEval-T}
\label{rq1}
\scriptsize
\setlength\tabcolsep{1.7pt}
\begin{threeparttable}
\begin{tabular}{lcccccccccccccccccc}
\toprule
\multirow{2.4}{*}{\textbf{Models}}                             & \multicolumn{3}{c}{\textbf{C++ to Python}}                                               & \multicolumn{3}{c}{\textbf{Python to C++}}                                               & \multicolumn{3}{c}{\textbf{Java to Python}}                                              & \multicolumn{3}{c}{\textbf{Python to Java}}                                              & \multicolumn{3}{c}{\textbf{C++ to Java}}                                                 & \multicolumn{3}{c}{\textbf{Java to C++}}                                                 \\ 
\cmidrule(lr){2-4} \cmidrule(lr){5-7} \cmidrule(lr){8-10} \cmidrule(lr){11-13} \cmidrule(lr){14-16}  \cmidrule(lr){17-19} 
                                  & \multicolumn{1}{l}{CA$_c$} & \multicolumn{1}{l}{CA$_m$} & \multicolumn{1}{l}{CSR} & \multicolumn{1}{l}{CA$_c$} & \multicolumn{1}{l}{CA$_m$} & \multicolumn{1}{l}{CSR} & \multicolumn{1}{l}{CA$_c$} & \multicolumn{1}{l}{CA$_m$} & \multicolumn{1}{l}{CSR} & \multicolumn{1}{l}{CA$_c$} & \multicolumn{1}{l}{CA$_m$} & \multicolumn{1}{l}{CSR} & \multicolumn{1}{l}{CA$_c$} & \multicolumn{1}{l}{CA$_m$} & \multicolumn{1}{l}{CSR} & \multicolumn{1}{l}{CA$_c$} & \multicolumn{1}{l}{CA$_m$} & \multicolumn{1}{l}{CSR} \\\midrule

Gemma & 19.15 & 36.44 & 89.01 & 1.06 & 2.94 & 9.57 & 12.41 & 31.43 & 78.72 & 4.26 & 5.35 & 18.79 & 3.90 & 5.09 & 26.24 & 5.32 & 5.01 & 9.93 \\
CodeLlama & 19.86 & 39.03 & 84.40 & 1.77 & 3.45 & 7.45 & 13.12 & 32.64 & 76.95 & 4.61 & 5.53 & 35.11 & 3.90 & 6.56 & 46.81 & 3.55 & 5.18 & 13.12 \\
CodeGemma & 25.89 & 48.70 & 90.78 & 6.38 & 8.12 & 25.89 & 18.09 & 41.62 & 83.33 & 10.99 & 13.30 & 34.40 & 7.45 & 11.14 & 56.74 & 8.87 & 8.89 & 24.11 \\
Llama3-8B & 27.30 & 50.86 & 89.01 & 5.32 & 6.74 & 32.98 & 18.09 & 46.63 & 89.72 & 5.32 & 7.60 & 44.33 & 6.38 & 10.88 & 61.70 & 11.35 & 11.14 & 33.69 \\
Llama3-70B & 35.46 & 58.55 & 90.07 & 6.38 & 6.91 & 32.98 & 25.89 & 55.44 & 94.68 & 9.22 & 11.14 & 71.28 & 15.6 & 18.74 & 77.66 & 11.35 & 10.71 & 30.14 \\

GPT-4o & 43.97 & 64.16 & 99.65 & 21.63 & 24.96 & 82.62 & 32.98 & 60.71 & \cellcolor[HTML]{E1FFFF}98.94 & 17.02 & 17.70 & 81.21 & 19.5 & 21.33 & \cellcolor[HTML]{E1FFFF}92.91 & 19.15 & 24.35 & 84.75 \\
DeepSeek& \cellcolor[HTML]{E1FFFF}45.04 & \cellcolor[HTML]{E1FFFF}65.98 & \cellcolor[HTML]{E1FFFF}100.0 & 24.47 & 29.79 & 80.5 & \cellcolor[HTML]{E1FFFF}35.11 & 60.97 & 98.58 & \cellcolor[HTML]{E1FFFF}26.60 & \cellcolor[HTML]{E1FFFF}27.37 & 87.59 & \cellcolor[HTML]{E1FFFF}22.34 & \cellcolor[HTML]{E1FFFF}24.18 & 89.36 & \cellcolor[HTML]{E1FFFF}20.92 & \cellcolor[HTML]{E1FFFF}24.53 & 78.72 \\

Claude & 42.20 & 65.98 & 99.65 & \cellcolor[HTML]{E1FFFF}27.30 & \cellcolor[HTML]{E1FFFF}30.92 & \cellcolor[HTML]{E1FFFF}85.82 & 31.56 & \cellcolor[HTML]{E1FFFF}61.49 & \cellcolor[HTML]{E1FFFF}98.94 & 20.21 & 22.63 & \cellcolor[HTML]{E1FFFF}87.94 & 14.54 & 16.15 & 91.13 & 19.86 & 22.11 &\cellcolor[HTML]{E1FFFF} 86.17 \\

Average & \textbf{32.36} & \textbf{53.71} & \textbf{92.82} & 11.79 & 14.23 & 44.73 & \underline{23.41} & \underline{48.87} & \underline{89.98} & 12.28 & 13.83 & 57.58& 11.70 & 14.26 & 67.82 & 12.55 & 13.99 & 45.08 \\
\bottomrule
\end{tabular}

\begin{tablenotes}
\definecolor{light cyan}{HTML}{E1FFFF}
\fboxsep1.5pt
\item[$\Phi$] \scriptsize Table cells with \colorbox{light cyan}{light cyan} background denote the highest performance in terms of CA$_c$/CA$_m$/CSR on each translation pair among various LLMs. Bold and underlined values in the Average row represent the highest and second-highest average performance respectively.
\end{tablenotes}
\end{threeparttable}
\vspace{-2em}
\end{table}

\textbf{Method-level code translation v.s. class-level code translation.} As shown in Figure \ref{comparison}, the code translation performance of LLMs degenerates significantly and consistently on all translation pairs when converting from method-level to class-level benchmark. {Specifically, all LLMs studied can achieve CA$_m$ scores over 60\% and CSR scores over 80\% on almost all translation pairs in Yang et al.'s benchmark. However, when assessing them on the ClassEval-T benchmark, LLMs' code translation performance declines 51.14\%--79.05\% on average in terms of CA$_m$ while 3.12\%--56.24\% on average in terms of CSR, where half of the translation results suffer over 80\% decline in terms of CA$_m$, and almost half of them suffer over 20\% decline in terms of CSR.} Besides, based on Yang et al.'s benchmark, it is hard to discriminate which LLM performs better, as method-level code translation benchmarks only contain standalone functions, they are shorter, lack diverse dependencies and complex contextual information, making LLMs easy to handle such samples and demonstrate inapparent performance discrepancies. Even in domain-specific PLs, such as Rust and Swift, a recent study \cite{tao2024unraveling} also unveiled a similar result. Therefore, we conclude that standalone functions are nearly no longer effective in assessing LLMs' code translation performance, but existing LLMs still have limited capability in complex coding tasks that bring about brand-new challenges, such as class-level code translation. 
\vspace{-0.3em}
\begin{boxK}
\small \faIcon{pencil-alt} \textbf{Finding 1:} Compared with method-level code translations, LLMs show a marked decline in performance when translating class-level code, and their performance discrepancies also appear with longer code, diverse dependencies, and complex contexts. 
\end{boxK}
\vspace{-0.3em} 

\textbf{Comparison among LLMs.} {Focusing on the results of the ClassEval-T benchmark, the three commercial LLMs consistently perform the best and are nearly neck-to-neck with each other, due to their much larger number of parameters than other LLMs \cite{abacha2024medec}. On average, commercial LLMs achieve a CA$_c$ score of 26.91\%, a CA$_m$ score of 36.96\%, and a CSR score of 90.25\% across various translation pairs.} Regarding the comparison between general and code LLMs, CodeGemma, a code LLM, always performs better than Gemma, a general LLM, on all evaluation metrics among various translation pairs.
However, Llama3-8B performs better than CodeLlama in most of the translation pairs. The reason behind this performance difference is that CodeGemma is enhanced from Gemma with an extra 500 billion code tokens for code-specific pretraining, resulting in its superior performance against Gemma. {In contrast, CodeLlama is based on Llama2 with an extra 500 billion code samples (1.25 times more than Llama2 \cite{touvron2023llama}) for finetuning, while Llama3-8B was pre-trained on 15 trillion tokens (seven times more than Llama2), with a relatively larger parameter size, leading to its superiority in code translation task against CodeLlama.}
The above phenomenon and analysis indicate that code LLMs are not necessarily better than general LLMs, even though they are specially pre-trained on code corpus. Delving deep into each category, CodeGemma constantly outperforms CodeLlama, while Llama3-8B constantly outperforms Gemma in terms of both CA and CSR, which are in accordance with evaluation results in other coding tasks conducted by recent studies \cite{team2024codegemma, dubey2024llama}. 
In addition, it is obvious that Llama3-70B always generates more samples that can pass test suites than Llama3-8B does among all translation pairs, even though Llama3-8B can generate more compilable samples in certain translation pairs, illustrating that LLMs with more parameters carry more powerful code translation ability, given other factors the same. 
\vspace{-0.3em}
\begin{boxK}
\small \faIcon{pencil-alt} \textbf{Finding 2:} 
{Compared with other LLMs, commercial LLMs exhibit predominately superior performance on class-level code translation.} Besides, Code LLMs do not necessarily outperform general LLMs in at least the class-level code translation task, and smaller LLMs normally perform worse, given other factors the same.  
\end{boxK}
\vspace{-0.3em}

\textbf{Comparison among PLs.} {Based on Figure \ref{comparison} and Table \ref{rq1}, LLMs generally perform much better on Python-oriented translations (e.g., C++/Java-to-Python) in terms of all evaluation metrics. However, concerning the rest of the code translation pairs, LLMs' performance drops dramatically. Specifically, for C++-oriented translations, the CA$_c$ decreased by 56.36\%, CA$_m$ decreased by 72.49\%, and CSR decreased by 50.87\%. For Java-oriented translations, each evaluation metric decreased by
57\%, 72.62\%, and 31.40\% in order.}
The performance differences can be attributed to pre-training data biases. Recent LLMs and those included in this paper normally use open-sourced code samples on GitHub for coding enhancement pretraining \cite{zhao2024deciphering, liu2024deepseek, team2024codegemma, roziere2023code, tao2024unraveling}, while Python has achieved the second most used PL on GitHub since 2022, as reported in \cite{Thetoppr42:online}. 
Therefore, LLMs potentially gained more knowledge of Python generation, which results in its much superior advantages against C++/Java-oriented translations.
\vspace{-0.3em}
\begin{boxK}
\small \faIcon{pencil-alt} \textbf{Finding 3:} 
Compared to C++/Java-oriented translation, LLMs manifest substantial superiority in translating other PLs to Python, owing to a relatively larger training base on Python-written code samples.
\end{boxK}
\vspace{-0.3em}

\subsection{RQ2: Translation Strategies}
\label{RQ2: Translation Strategies}
\begin{figure}[htbp]
\vspace{-1em}
\setlength{\abovecaptionskip}{0cm}
\includegraphics[width=0.7\textwidth]
{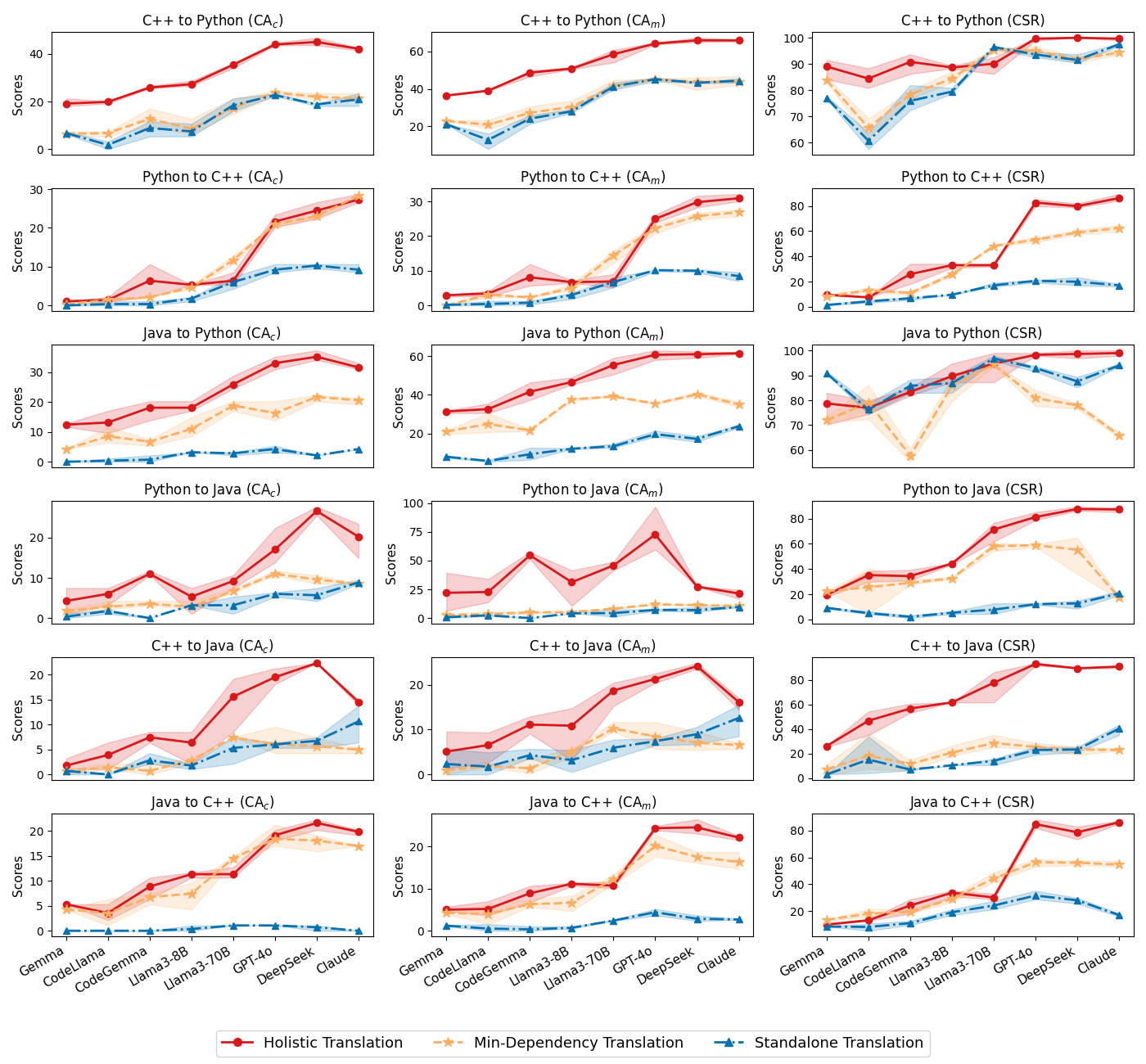}
\caption{Translation results in terms of CA and CSR on ClassEval-T with different strategies}
\label{rq2}
\vspace{-1em}
\end{figure}

Figure \ref{rq2} demonstrates the experimental results on ClassEval-T with different code translation strategies mentioned in Section \ref{Studied Translation Strategies}, where each column showcases experimental results of diverse LLMs in terms of each evaluation metric (either CA$_c$, CA$_m$, or CSR) among six translation pairs. All are recorded in their average scores wrapped with ranges among repeated experiments. Our observations are listed below. 

\textbf{Analysis among LLMs.}
{As shown in Figure \ref{rq2}, the holistic translation strategy is always the best selection for commercial LLMs, outperforming the other two strategies by an average of 341.51\% in terms of CA$_c$, by 310.62\% in terms of CA$_m$, and by 157.64\% in terms of CSR, while for other LLMs, it depends on different translation scenarios. 
One potential reason might be that commercial LLMs carry a large number of parameters (>100B), which enhances their ability to understand and generate code effectively. Thus, it is not difficult for these models to complete the code translation task given the entire class as inputs in the holistic strategy. However, other LLMs, with smaller parameter sizes, carry relatively limited capability in leveraging long input contexts, as revealed by \cite{liu2024lost}. On the other hand, we observed that, on average, 8.87\% of the test samples exceeded the model’s context window length limitation, which in turn constrains the applicability of the holistic translation strategy to some extent. The specific context window exceeding rates for each model are detailed in the repository's table\cite{wLinHooC78:online}. As a result, holistic translation strategy does not show overwhelming advantages against min-dependecy strategy, and in some cases, the min-dependency strategy does indeed yield higher performance.}
Besides, LLMs with the standalone strategy perform the worst on CA$_c$ and CA$_m$ in most of the translation scenarios, as we do not provide contextual dependencies but a standalone class skeleton or method during the translation, leading to the translated methods or fields being highly possible to have conflicts with other translated counterparts in classes. For example, one method invoked by another may be translated to different method names during the translation of its caller and itself, causing conflicts in a class. However, this is not the case in terms of CSR. When translating with the standalone strategy, LLMs tend to generate some contextual dependencies based on their imagination; thus, discriminating the translated results from other generated contexts via manually crafted rules is one of our processes for the standalone translation strategy. Considering that the translation process of LLMs is uncontrollable, which means they may rename the source program completely, making our rules difficult to capture the translated results. In this paper, we deem the above a failed translation, and no method will be inserted into the translated class skeleton. This will definitely affect functionality correctness but not the program compilation, leading to an overestimation of CSR among LLMs in the standalone translation strategy. 
\vspace{-0.3em}
\begin{boxK}
\small \faIcon{pencil-alt} \textbf{Finding 4:} 
{Commercial LLMs always work much better on the holistic translation strategy owing to their powerful code understanding and generation ability.} However, the selection of translation strategy for smaller LLMs depends on different scenarios. The standalone strategy performs the worst for most translation scenarios. 
\end{boxK}
\vspace{-0.3em}

\textbf{Analysis among PLs.}
{The selection of translation strategy also relies on target PLs. As shown in Figure \ref{rq2}, the holistic strategy always performs the best in Python-oriented and Java-oriented translations in terms of CA$_c$ and CA$_m$, outperforming other strategies on average by 42.05\%--84.40\% in terms of CA$_c$ while by 35.22\%--59.09\% in terms of CA$_m$. However, focusing on C++-oriented translation, most LLMs with holistic and min-dependency strategies perform neck-to-neck, except for commercial LLMs whose holistic strategy still outperforms but the gap to min-dependecy strategy shrinks. The reasons are multifaceted: (1) Because C++ requires explicit memory allocation and deallocation, lacks rich third-party libraries, and has other language-specific characteristics, implementing the same functionality typically results in longer programs compared to Java and Python. Based on tokenization averaging across all LLMs, C++ code in ClassEval-T has an average of 736.91 tokens, while Python and Java code contains an average of 332.29 and 575.15 tokens.   Commercial LLMs have large input and output context windows, so no samples are truncated due to token limitations, allowing the holistic strategy to remain the most optimal.  However, for other LLMs, although only an average of 1.43\% of samples exceed the LLM’s input limit, a larger challenge arises with C++ data. An expected 19.79\% of C++ samples exceeds the output token limit, compared to only 0.64\% for Python and 6.17\% for Java (To simplify the statistical process, we use the ground truth as the estimated output length). This means that, while most samples can be fully input into the LLM, a significantly higher proportion of C++ samples fail to output completely, reducing the effectiveness of the holistic strategy.  As a result, when translating C++, the min-dependency and holistic strategies perform neck-and-neck for LLMs other than commercial LLMs.}(2) The min-dependency strategy aligns with C++'s separate compilation feature, enabling flexible code splicing and tolerance for unimplemented sections \cite{favre1995cpp}.  In addition, C++ is more complex than Python and Java, especially in terms of manual memory management, template systems, and strict dependency management. Generating code step by step reduces the information that LLM needs to pay attention to each time, reducing the possibility of LLM generating incorrect code. Therefore, in the case of translating other PLs to C++, the gap between the holistic and min-dependency strategies is smaller for commercial LLMs.
\vspace{-0.3em}
\begin{boxK}
\small \faIcon{pencil-alt} \textbf{Finding 5:} 
{Python-oriented and Java-oriented translations work better on the holistic strategy. C++-oriented translations perform better on the holistic strategy in commercial models, while in other models, holistic strategy and min-dependency strategy show neck-to-neck performance.}
\end{boxK}
\vspace{-0.3em}

\begin{figure}[htbp]
\vspace{-1.5em}
\setlength{\abovecaptionskip}{0cm}
\includegraphics[width=1\textwidth]
{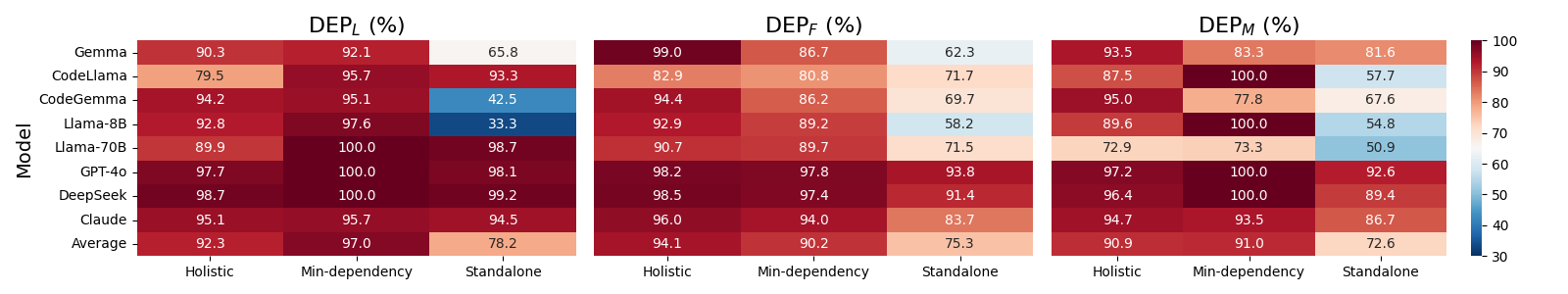}
\caption{Dependency awareness analysis among various LLMs}
\label{rq3}
\vspace{-1.5em}
\end{figure}

\subsection{RQ3: Dependency Awareness}
\label{RQ3: Dependency Awareness}

The code translation process of LLMs is uncontrollable to some extent, which means it is hard to restrict how LLMs define or name each translated field, variable, and method strictly. Even worse, we also cannot forecast which library LLM will import to implement the equivalent functionality of source PLs. Therefore, it is hard to automatically measure to what extent can LLMs be aware of the necessary dependencies. As such, we conduct human evaluation as a compromise. {Specifically, for each translation strategy, we randomly select 20 class-level samples in each LLM's translation results, a total of 480 selected samples.} Afterward, six authors of this paper with over three years of coding experience in Python, Java, and C++ are involved as evaluators and are uniformly divided into three groups. Each group is assigned to one PL-oriented translation for analysis. For the measurement of $d_{F}(\cdot)$ and $d_{M}(\cdot)$ mentioned in Section \ref{Evaluation Metrics}, evaluators are required to manually count the number of fields and methods that each caller method relies on. For the measurement of $d_{L}(\cdot)$, we consider three cases as an effective dependency awareness: (1) A translated program invokes the one-to-one-mapping library of the source PL, (2) A translated program invokes one-to-many- or many-to-one-mapping libraries of the source PL, (3) A translated program need not invoke any library, but still implements the equivalent functionality of the source PL. After counting the number of valid dependencies, we compute DEP$_F$, DEP$_M$, and DEP$_L$, respectively, and the corresponding results are shown in Figure \ref{rq3}. 

{As can be seen, LLMs with the holistic translation strategy perform the best on field dependency awareness, outperforming other strategies by 4.00\%--24.97\% on average, while LLMs with the min-dependency strategy work much better on invoking useful library dependencies, surpassing others by 5.09\%--24.04\% on average. As for method dependency awareness, LLMs with holistic and min-dependency strategies perform neck-to-neck, both achieving a DEP$_M$ score of around 91\% on average.} However, LLMs with the standalone translation strategy always perform the worst. A potential explanation is fields need more contextual information to make LLMs observe how different methods exert them to understand their utility, while methods with their implementations to facilitate understanding need relatively less contextual information compared with fields. Therefore, the holistic strategy dominates in DEP$_F$ but performs neck-to-neck with the min-dependency strategy in DEP$_M$. As for library dependencies, they are more difficult for LLMs to infer as there are no explicit hints from the source PL, and need a deeper understanding of the program functionalities under translation. With the min-dependency strategy, LLMs can focus more on partial programs, while the holistic strategy brings more noise. {In addition, among all LLMs, the commercial LLMs still outperform other LLMs in almost all dependency evaluations and translation strategies, surpassing others on average by 16.19\%, 15.64\%, and 19.58\% in terms of DEP$_F$, DEP$_M$, and DEP$_L$, respectively.}
\vspace{-0.5em}
\begin{boxK}
\small \faIcon{pencil-alt} \textbf{Finding 6:} 
The holistic translation strategy improves LLMs' awareness of field dependencies, while the min-dependency strategy allows LLMs to invoke more necessary libraries. As for method dependencies, the above two strategies perform neck-to-neck. 
\end{boxK}
\vspace{-0.5em}

\vspace{-0.8em}
\subsection{RQ4: Failed Cases Analysis}

{As one of the top-performing commercial models,  Deepseek serves as a valuable benchmark for understanding the limitations of state-of-the-art code translation systems. } By analyzing its failures, we aim to identify persistent challenges in code translation, highlighting areas where even the most advanced models require further refinement.
{Therefore, we dedicate 240 person-hours to thoroughly evaluate Deepseek’s 1,243 error cases with the thematic analysis methodology \cite{cruzes2011recommended}, where the error cases are selected from a single-time experiment.} First, we engage three experts (co-authors in this paper), each with over five years of development experience in Python, Java, and C++.  These experts analyze 20\% samples of the error cases to identify and summarize the underlying causes.  Following this initial analysis, the experts hold a detailed discussion to organize the identified errors into categories, establishing specific criteria and examples for each type.  These guidelines are compiled into a codebook, ensuring uniformity and clarity in subsequent evaluations.
{Subsequently, we recruit eight evaluators with at least 3 years of experience in Python, Java, and C++. All evaluators are from computer science backgrounds and have achieved scores of no lower than an A in programming courses related to these languages.}
 Each sample is assessed through a rigorous, double-blind, dual-review process, where each evaluator independently examine the sample without knowledge of the other’s findings.  If both evaluators agree on the error classification, their result is recorded as final.  In cases where evaluations diverge, evaluators engage in a structured discussion to reach a consensus on the error type and classification.  When agreement could not be reached, the case is escalated to the expert team for arbitration, ensuring consistent and accurate categorization.  To streamline the evaluation process, we develop a small tool using Python’s tkinter library \cite{tkinter—69:online} to assist evaluators in recording results efficiently.  This tool, which we have open-sourced in our repository, allows evaluators to log and categorize errors quickly, reducing manual effort and enhancing overall evaluation efficiency.  Throughout the process, if an evaluator identify an error type not covered in the codebook, they are encouraged to report it to the expert team, who would then review and, if necessary, update the codebook to accommodate the new error category.
Considering that each sample could contain multiple errors, we categorize on statement/token-level granularity, ensuring the exhaustiveness of the categorization of samples.
For ambiguous errors, evaluators discuss to determine the most appropriate classification.

\begin{table}[]
\vspace{-1.5em}
\setlength{\abovecaptionskip}{0cm}
\caption{Error types and frequencies across different translation pairs and strategies}
\label{tab:error_types}
\scriptsize
\setlength\tabcolsep{3pt}
\begin{threeparttable}
\resizebox{\linewidth}{!}{ 
\begin{tabular}{lcccccccccccccccccc}
\toprule
\multicolumn{1}{r}{\textbf{Strategy}} & \multicolumn{6}{c}{\textbf{Holistic}} & \multicolumn{6}{c}{\textbf{Min-Dependency}} & \multicolumn{6}{c}{\textbf{Standalone}} \\
\cmidrule(lr){2-7} \cmidrule(lr){8-13} \cmidrule(lr){14-19}

\multicolumn{1}{r}{\textbf{Error Type (\%)\hspace{26pt}Target PL}} & \multicolumn{2}{c}{\textbf{C++}} & \multicolumn{2}{c}{\textbf{Python}} & \multicolumn{2}{c}{\textbf{Java}} & 
                                              \multicolumn{2}{c}{\textbf{C++}} & \multicolumn{2}{c}{\textbf{Python}} & \multicolumn{2}{c}{\textbf{Java}} & 
                                              \multicolumn{2}{c}{\textbf{C++}} & \multicolumn{2}{c}{\textbf{Python}} & \multicolumn{2}{c}{\textbf{Java}} \\
\cmidrule(lr){2-3} \cmidrule(lr){4-5} \cmidrule(lr){6-7} \cmidrule(lr){8-9} \cmidrule(lr){10-11} \cmidrule(lr){12-13} \cmidrule(lr){14-15} \cmidrule(lr){16-17} \cmidrule(lr){18-19}

\multicolumn{1}{r}{\textbf{Source PL}} & \textbf{Java} & \textbf{Py} & \textbf{Java} & \textbf{C++} & \textbf{Py} & \textbf{C++} & 
                                              \textbf{Java} & \textbf{Py} & \textbf{Java} & \textbf{C++} & \textbf{Py} & \textbf{C++} & 
                                              \textbf{Java} & \textbf{Py} & \textbf{Java} & \textbf{C++} & \textbf{Py} & \textbf{C++} \\
\midrule

        \rowcolor{blue!35} \textbf{Library-related} & 9.88 & 6.90 & 0.00 & 0.00 & 12.50 & 10.71 & 13.00 & 9.30 & 22.22 & 8.33 & 13.10 & 17.27 & 10.24 & 7.92 & 9.78 & 8.86 & 20.71 & 24.19 \\
        \rowcolor{blue!15} Missing required library & 7.41 & 5.17 & 0.00 & 0.00 & 8.33 & 10.71 & 8.00 & 6.98 & 22.22 & 6.94 & 13.10 & 17.27 & 7.09 & 5.94 & 9.78 & 8.86 & 17.75 & 20.97 \\
        \rowcolor{blue!15} API/library mismatch & 2.47 & 1.72 & 0.00 & 0.00 & 4.17 & 0.00 & 5.00 & 1.16 & 0.00 & 1.39 & 0.00 & 0.00 & 1.57 & 0.99 & 0.00 & 0.00 & 2.96 & 3.23 \\
        \rowcolor{blue!15} Copying libraries from source PL & 0.00 & 0.00 & 0.00 & 0.00 & 0.00 & 0.00 & 0.00 & 1.16& 0.00 & 0.00 & 0.00 & 0.00& 1.57 & 0.99 & 0.00 & 0.00& 0.00 & 0.00\\
        \hline
\rowcolor{green!45} \textbf{Syntax errors} & 30.86 & \textbf{37.93} & 3.23 & 0.00 & \textbf{33.33} & \textbf{46.43} & \textbf{45.00} & \textbf{40.70} & 2.78 & 4.17 & \textbf{39.29} & 28.18 & \textbf{43.31} & \textbf{47.52} & 1.09 & 1.27 & 28.40 & 28.23 \\
\rowcolor{green!20} Symbol errors & 0.00 & 0.00 & 0.00 & 0.00 & 8.33 & 7.14 & 5.00 & 5.81 & 0.00 & 0.00 & 11.90 & 6.36 & 4.72 & 5.94 & 0.00 & 0.00 & 0.00 & 8.06 \\
\rowcolor{green!20} Language feature errors & 20.99 & 24.14 & 0.00 & 0.00 & 4.17 & 25.00 & 22.00 & 20.93 & 1.39 & 4.17 & 13.10 & 8.18 & 18.90 & 24.75 & 1.09 & 0.00 & 21.30 & 8.87 \\
\rowcolor{green!20} Type mismatch errors & 8.64 & 13.79 & 0.00 & 0.00 & 8.33 & 14.29 & 12.00 & 13.95 & 0.00 & 0.00 & 9.52 & 11.82 & 17.32 & 14.85 & 0.00 & 0.00 & 3.55 & 7.26 \\
\rowcolor{green!20} Grammar structure errors & 0.00 & 0.00 & 3.23 & 0.00 & 8.33 & 0.00 & 3.00 & 0.00 & 0.00 & 0.00 & 3.57 & 1.82 & 0.00 & 0.00 & 0.00 & 0.00 & 2.37 & 2.42 \\
\rowcolor{green!20} Retaining source PL syntax & 1.23 & 0.00 & 0.00 & 0.00 & 4.17 & 0.00 & 3.00 & 0.00 & 1.39 & 0.00 & 1.19 & 0.00 & 2.36 & 1.98 & 0.00 & 1.27 & 1.18 & 1.61 \\
        
        \hline
        \rowcolor{purple!40} \textbf{Function/variable usage issues} & \textbf{41.98} & 17.24 & 1.61 & 0.00 & 25.00 & 21.43 & 29.00 & 27.91 & 2.78 & 0.00 & 28.57 & \textbf{31.82} & 33.07 & 30.69 & 1.09 & 2.53 & \textbf{36.09} & \textbf{38.71} \\
        \rowcolor{purple!20} Functions missing return values & 0.00 & 0.00 & 0.00 & 0.00 & 4.17 & 3.57 & 0.00 & 0.00 & 0.00 & 0.00 & 1.19 & 0.00 & 0.00 & 0.00 & 0.00 & 0.00 & 2.37 & 2.42 \\
        \rowcolor{purple!20} Return value type mismatch & 11.11 & 0.00 & 0.00 & 0.00 & 4.17 & 7.14 & 4.00 & 8.14 & 0.00 & 0.00 & 3.57 & 6.36 & 3.15 & 0.99 & 0.00 & 0.00 & 4.14 & 3.23 \\
        \rowcolor{purple!20} Attempt to call non-existent function & 8.64 & 1.72 & 0.00 & 0.00 & 4.17 & 7.14 & 8.00 & 3.49 & 0.00 & 0.00 & 9.52 & 10.00 & 9.45 & 2.97 & 0.00 & 0.00 & 17.75 & 20.16 \\
        \rowcolor{purple!20} Attempt to call undefined fields & 3.70 & 3.45 & 1.61 & 0.00 & 8.33 & 0.00 & 10.00 & 11.63 & 2.78 & 0.00 & 14.29 & 13.64 & 14.17 & 14.85 & 1.09 & 1.27 & 10.65 & 12.90 \\
        \rowcolor{purple!20} Parameter errors & 18.52 & 12.07 & 0.00 & 0.00 & 4.17 & 3.57 & 7.00 & 4.65 & 0.00 & 0.00 & 0.00 & 1.82 & 6.30 & 11.88 & 0.00 & 1.27 & 1.18 & 0.00 \\
        
        \hline
        \rowcolor{orange!40} \textbf{Code consistency/completeness} & 3.70 & 6.90 & 0.00 & 0.00 & 8.33 & 0.00 & 4.00 & 1.16 & 1.39 & 0.00 & 5.95 & 16.36 & 3.94 & 8.91 & 0.00 & 1.27 & 5.33 & 2.42 \\
        \rowcolor{orange!20} Generated code inconsistency & 3.70 & 1.72 & 0.00 & 0.00 & 8.33 & 0.00 & 4.00 & 1.16 & 1.39 & 0.00 & 5.95 & 13.64 & 3.94 & 7.92 & 0.00 & 1.27 & 1.18 & 0.81 \\
        \rowcolor{orange!20} Non-code content generated & 0.00 & 0.00 & 0.00 & 0.00 & 0.00 & 0.00 & 0.00 & 2.73 & 0.00 & 0.00 & 0.00 & 3.28 & 0.00 & 0.00 & 0.00 & 0.00 & 0.59 & 1.61 \\
        \rowcolor{orange!20} Generated code incomplete & 0.00 & 5.17 & 0.00 & 0.00 & 0.00 & 0.00 & 0.00 & 0.00 & 0.00 & 0.00 & 0.00 & 0.00 & 0.00 & 0.99 & 0.00 & 0.00 & 3.55 & 0.00 \\
        
        \hline
        \rowcolor{brown!40} \textbf{Functional errors} & 6.17 & 22.41 & \textbf{95.16} & \textbf{100.00}& 12.50 & 17.86 & 2.00 & 12.79 & \textbf{70.83} & \textbf{87.50} & 8.33 & 3.64 & 4.72 & 0.99 & \textbf{88.04} & \textbf{86.08} & 8.28 & 3.23 \\
        \hline
        \rowcolor{yellow!40} \textbf{Runtime errors} & 0.00 & 0.00 & 0.00 & 0.00 & 4.17 & 0.00 & 0.00 & 1.16 & 0.00 & 0.00 & 0.00 & 0.00 & 0.00 & 1.57 & 0.00 & 0.00 & 0.00 & 0.00 \\

        \hline
        \rowcolor{gray!30} \textbf{Other errors} & 7.41 & 8.62 & 0.00 & 0.00 & 4.17 & 3.57 & 7.00 & 6.98 & 0.00 & 0.00 & 4.76 & 2.73 & 3.15 & 3.96 & 0.00 & 0.00 & 1.18 & 3.23 \\
\bottomrule
\end{tabular}}
\end{threeparttable}
\vspace{-2em}
\end{table}

Table \ref{tab:error_types} presents the evaluation results for three translation strategies, namely holistic, min-dependency, and standalone, across ClassEval-T, categorized by target and source PLs. Major error types include library-related errors, syntax errors, function/variable usage issues, code consistency/completeness errors, generated code functional errors, runtime errors, and Other errors, with each type further divided into specific error types. The results reveal distinct patterns in translation errors based on target and source PL pairings.
(1) {Functional errors are especially prominent in Python-oriented translations, accounting for 87.94\% across all three translation strategies on average.} Considering the outstanding performance of DeepSeek in Python-oriented translations mentioned in RQ1, we conclude that other errors in DeepSeek's Python-oriented translations have been almost eliminated owing to its exceptional capability in code understanding and generation while generating semantic equivalent programs has become its primary challenge. 
(2) {Syntax errors, such as symbol and language feature issues, frequently occur in C++/Java-oriented translations across different strategies, accounting for 38.25\% and 33.98\%, respectively, which is also revealed by a recent method-level code translation study \cite{yang2024exploring}.} This pattern can be attributed to the strict syntax and type requirements of these PLs, which pose challenges when translating from more flexible PLs, such as Python. The structural differences between C++ and Java further compound syntax issues when translating between the two.
(3) {Function and variable usage issues, often arising within class structures, frequently occur in C++/Java-oriented translations across various strategies, constituting 33.39\% and 30.27\% of the total errors, respectively.}
These issues, along with code consistency errors, are especially prevalent in the standalone strategy, where separate translation and missing contextual dependencies limit the LLM’s ability to identify function and field references, increasing the likelihood of calling non-existent functions or fields. Additionally, merging independently translated segments without contextual information often results in logical inconsistencies.
To provide a clearer understanding of these types, the following section presents examples illustrating each error type and describes what each entails in practical terms.

\begin{figure}[htbp]
\vspace{-1em}
\setlength{\abovecaptionskip}{0cm}
\includegraphics[width=0.7\textwidth]
{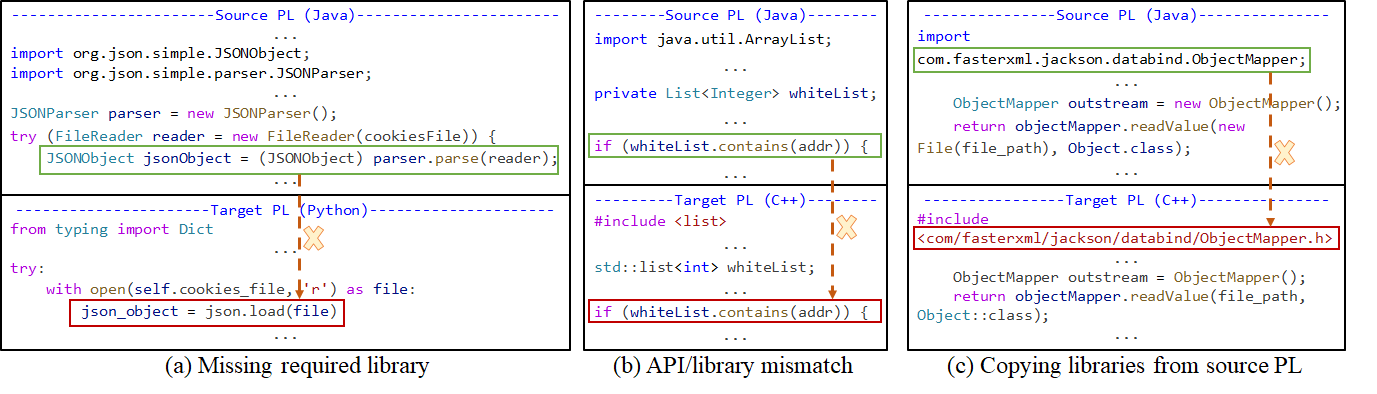}
\caption{Failed examples for Library-related errors}
\label{rq4-1}
\vspace{-1em}
\end{figure}

\textbf{Library-related.}
(1) Missing required library: This category means essential modules or dependencies are omitted during translation. {As shown in the example in Figure {\ref{rq4-1}} (a), the Java code snippet aims to parse a \textit{FileReader} object to \textit{JSONObject} one. After the translation to Python code, DeepSeek omits to add the import statement for the corresponding library, i.e., \textit{json}, but directly uses it, as specified in the red box.} 
In RQ3, we find that the min-dependency approach had a high $DEP_L$, and here it still results in some missing library errors. This is because, during integration, we include only the translated class skeletons' libraries and fields, while methods are integrated with their bodies only. Although losing some separately translated dependencies, this solution prevents library conflicts, reduces logical confusion, and simplifies integration. (2) API/library mismatch: This issue occurs when library-specific methods from the source PL are incorrectly applied to the target PL’s libraries. In Figure \ref{rq4-1} (b), a Java method, \textit{contains}, in \textit{WhiteList} is copied directly to C++'s \textit{std::list}, resulting in an error, as \textit{std::list} does not support the method, \textit{contains}, by default. (3) Copying libraries from source PL: This category includes errors where libraries from the source PL are directly copied into the target PL without proper adaptation. For example (as shown in Figure \ref{rq4-1} (c)), Java’s \textit{com.fasterxml.jackson.databind.ObjectMapper} is directly translated to C++ as \textit{\#include <com/fasterxml/jackson/databind/ObjectMapper.h>}, which does not exist in C++.

\begin{figure}[htbp]
\vspace{-1em}
\setlength{\abovecaptionskip}{0cm}
\includegraphics[width=0.7\textwidth]
{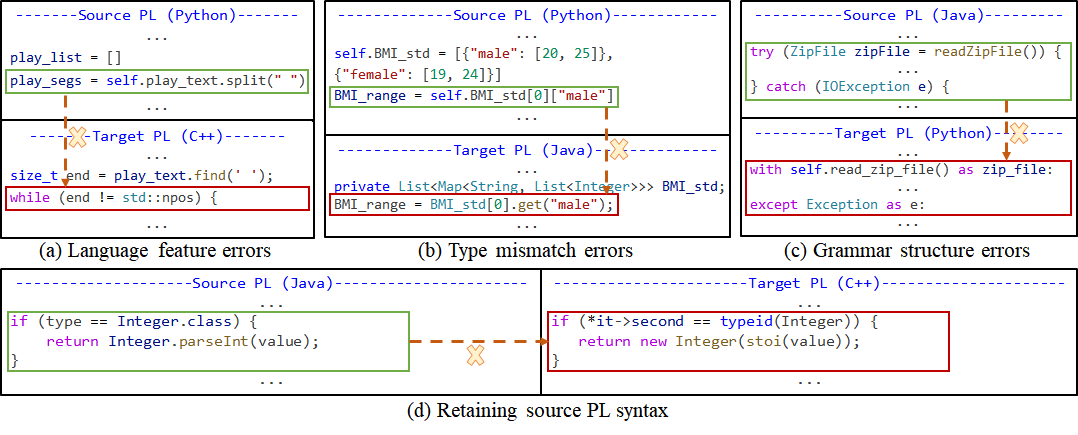}
\caption{Failed examples for Syntax errors}
\label{rq4-2}
\vspace{-1.2em}
\end{figure}

\textbf{Syntax errors.} (1) Symbol errors: Symbol errors occur when symbols are missing or misplaced, such as missing parentheses ``\textit{\{}'' or semicolons ``\textit{;}'', which are required in languages like C++ but not in Python. These missing symbols lead to syntax disruptions when translating between PLs with differing symbol requirements. (2) Language feature errors: These errors arise from misunderstandings of language-specific conventions. {For example, as shown in Figure {\ref{rq4-2}} (a), when implementing Python's \textit{split} method in C++, the incorrect statement \textit{while (end! = std::npos)} is used instead of the correct statement \textit{std::string::npos}.} 
(3) Type mismatch errors: Type mismatch errors involve discrepancies in expected types, which lead to incompatible or undefined behavior. Figure \ref{rq4-2}  (b) presents an example where a \textit{HashMap<Object, Integer>} is used instead of the expected \textit{HashMap<String, Integer>}, causing type incompatibility in the target PL. (4) Grammar structure errors: {occur when incorrect syntax structures are used. As shown in Figure \ref{rq4-2} (c), the translated Python code attempts to capture an exception with only the code block, \textit{except Exception as e:}, but ignores the block, \textit{try}, disrupting the logical flow.} (5) Retaining source PL syntax: This issue occurs when source PL syntax is translated directly into the target PL. {For example, as shown in Figure {\ref{rq4-2}} (d), \textit{Integer} is a data structure exclusively used in Java code but is directly copied to C++ code.}

\begin{figure}[htbp]
\vspace{-0.5em}
\setlength{\abovecaptionskip}{0cm}
\includegraphics[width=0.7\textwidth]
{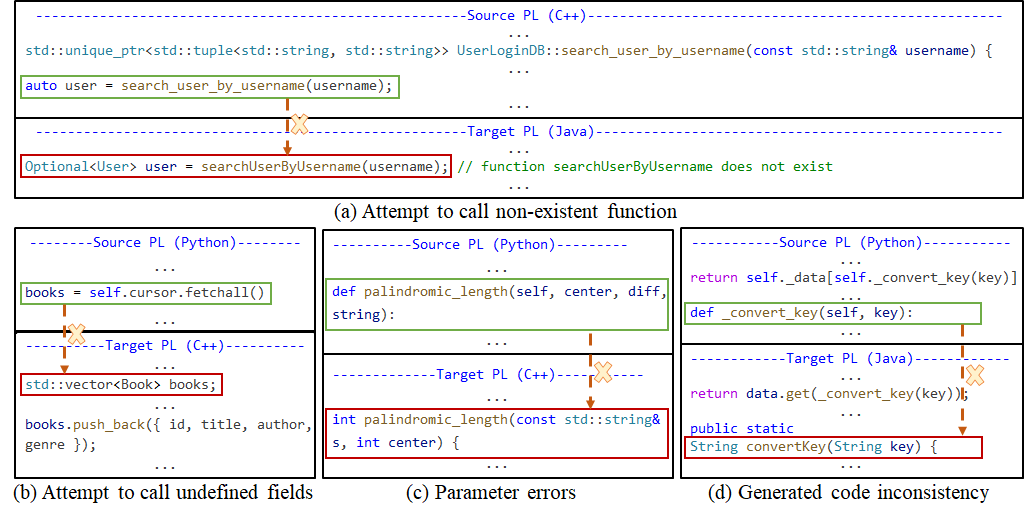}
\caption{Failed examples for Function/variable usage issues and Generated code inconsistency}
\label{rq4-3}
\vspace{-0.9em}
\end{figure}

\textbf{Function/variable usage issues.} (1) Functions missing return values: Errors in this type occur when a function fails to provide a required return value. This leads to incomplete execution of intended logic, causing downstream errors when the function’s output is used in subsequent operations. (2) Return value type mismatch: This error happens when the return type of a function does not align with its expected type. (3) Attempt to call non-existent functions: This issue arises when the target code attempts to call a function that has not been defined or implemented. For example, as shown in Figure \ref{rq4-3} (a), the source function \textit{search\_user\_by\_username} is translated to the target code as \textit{searchUserByUsername}, but the function \textit{searchUserByUsername} is not implemented. (4) Attempt to call undefined fields: This error occurs when variables or fields are used in the target code without being defined or declared. For example, as shown in Figure \ref{rq4-3} (b), the field \textit{books} is initialized as a \textit{std::vector<Book>} and used without prior declaration, causing undefined behavior. (5) Parameter errors: Parameter errors include incorrect parameter types, numbers, or usage in function calls. {As shown in Figure {\ref{rq4-3}} (c), an error occurs when the function \textit{palindromic\_length} is called with two parameters in C++ instead of the required 3 parameters, resulting in incorrect function usage.}

\textbf{Code consistency/completeness.} (1) Generated code inconsistency: it occurs when the same variable or function is referenced inconsistently within the same code context, leading to logical errors and a lack of coherence. {For example, as shown in Figure {\ref{rq4-3}} (d), the Python function \textit{\_convert\_key} is translated inconsistently in Java within the same code block: it appears as \textit{\_convert\_key} in one line and as \textit{convertKey} in another, resulting in compilation errors due to mismatched function names.} (2) Non-code content generated: Non-code content refers to instances where irrelevant information, such as \textit{``Octal:15''} or other unrelated descriptions, is produced instead of code. This also includes cases where the LLM refuses to translate a segment entirely. (3) Generated code incomplete: it describes translations where, due to limitations in model capability or token constraints, essential parts of the code are missing or forcibly truncated before completion, resulting in an incomplete program.

\textbf{Functional and Runtime errors.} Functional errors occur when the translated code fails to replicate the intended functionality of the source code.  These errors stem from misinterpretations in logic, structure, or data handling, resulting in outputs that deviate from the expected behavior in the target PL.  Although the translated code is syntactically correct, functional errors disrupt the code’s ability to perform as intended.
Runtime errors are issues that arise during the execution of translated code, leading to unexpected interruptions or failures. These errors prevent the program from running smoothly in the target environment and may result in crashes, incorrect outputs, or halted processes.
\vspace{-0.5em}
\begin{boxK}
\small \faIcon{pencil-alt} \textbf{Finding 7:} 
Functional errors are most common in Python-oriented translations, as other errors are almost resolved for DeepSeek. Similar to those in method-level translations, syntax errors remain a primary issue in class-level translations. In C++/Java-oriented translations, class-related errors, such as function/variable usage and consistency issues, are more prominent.
\end{boxK}
\vspace{-1em}

\section{Implications}
This paper serves as the first class-level code translation benchmark, unravelling more deficiencies of recent LLMs that are not exposed in previous statement- and method-level code translation benchmarks. Based on our findings during the whole experiment, we summarize the following implications.

\textbf{Implications for researchers:} 
As the first study focusing on class-level code translation, this paper revealed the defects of previous method-level code translation benchmarks and exposed LLMs' inability to translate code for an entire class in RQ1.  In particular, our contributed benchmark with class-level code, diverse dependencies, and alignment with practical development significantly fosters the code translation research to the era of class level. {
While repository-level code translation remains a long-term objective, significant challenges arise from cross-file dependencies and the context window limitations of LLMs, making complete translation in a single pass difficult.  However, experimental findings indicate that holistic translation is optimal when the context window is sufficiently large.  Therefore, future repo-level research should aim to maximize granularity in code splitting to leverage larger context windows effectively.  This study suggests that, for subsequent repo-level investigations, adopting larger code units such as classes or modules can significantly improve translation quality, providing a crucial stepping stone towards achieving comprehensive repository-level translations. Our findings indicate that, even at class-level granularity, LLMs struggle with structural integrity, dependency handling, and code correctness, highlighting the necessity of a phased approach.  By first addressing the challenges at the class level, our work provides a solid foundation, offering a finer-grained perspective that positions future research to progressively and efficiently tackle real-world translation scenarios.}
In addition, RQ4 unveiled a series of failure cases that the best-performing LLM made, enlightening future research constantly focusing on resolving syntax, functional, and runtime errors that LLMs also suffer in method-level code translations, but also pay attention to those errors mainly occur in class-level code translations, such as library-related, function/variable usage, and code consistency/completeness issues.  

\textbf{Implications for practitioners:}
Based on our contributed benchmark, namely ClassEval-T, we conducted extensive experiments and concluded insightful findings, which can guide developers' practical usage of LLMs for class-level code translation. For example, RQ2-3 uncovered the discrepancies in correctness and dependency awareness among diverse translation strategies, LLMs, and PLs, suggesting that powerful LLMs, such as DeepSeek, and Python/Java-oriented translations are more suitable for holistic strategy; The selection of translation strategy for smaller LLMs depends on different scenarios, and for C++-oriented translations, the min-dependency strategy and holistic approach are neck-to-neck in terms of performance. Therefore, choosing a suitable translation strategy can boost the class-level code translation performance. Besides, RQ4 also reported and categorized diverse errors the best-performing LLMs may make, precisely reminding practitioners to notice their potential risks when deployment.

\section{Threats to Validity}
\textbf{Threats to external validity} relates to the generality of our experimental findings and failure taxonomy \cite{zhang2024hard,liu2024exploring}. Our newly constructed benchmark specifically targets code translation among Python, Java, and C++ without the consideration of many other domain-specific PLs. Nonetheless, ClassEval-T contains more practical coding tasks on the class level than any other previous counterparts, making our findings and failures of LLMs more useful and valuable for daily development. Besides, Python, Java, and C++ are the most widely studied and used PLs, which are much more meaningful for exploration. 
As for failure taxonomy, we thoroughly evaluated {1,243} class-level error cases made by DeepSeek across all three translation strategies. Besides, a thematic-analysis-driven methodology is strictly followed, ensuring the generality and correctness of our conclusion. Therefore, we believe the above threats are minimal.       

\noindent\textbf{Threats to internal validity} lies in the potential of data leakage and processing involving manual handling (e.g., benchmark manual construction and manual evaluation) \cite{ouyang2024benchmarking,xia2023automated}. 
ClassEval-T is a manually constructed code translation benchmark extended from ClassEval with extra parallel Java and C++ programs. It is inevitable that Python programs in ClassEval may have been crawled by our studied LLMs for training, but their corresponding Java and C++ programs are unseen to LLMs. No matter which translation pairs are experimented with, at least code samples on one side (source or target) are assured unseen to LLMs. In addition, 0\% of translated samples exactly match the ground truths. Therefore, the data leakage threat is extremely limited. {A related concern is the potential future contamination of open-source benchmarks as LLMs evolve and may eventually be trained on ClassEval-T. To mitigate this, we intend to periodically update the dataset using program mutation rules that systematically transform benchmark programs while preserving functional equivalence. This approach maintains benchmark reliability, ensuring its continued relevance and reducing the risk of LLMs memorizing specific samples.} Additionally, to guarantee the quality of our constructed benchmark, we predefine rigorous translation principles and include multiple participators to alleviate the subjectiveness and mistakes in manual coding. As for the accuracy of manual evaluation in dependency analysis (RQ3) and failure analysis (RQ4), multiple experienced participators are also included for assessment. Moreover, we follow the methodology of thematic analysis with a detailed codebook for guidelines and a rigorous, double-blind, and dual-review process in failure categorization, thereby minimizing this internal threat.

\noindent\textbf{Threats to construct validity} concerns the comprehensiveness of evaluation metrics \cite{yang2021multi,dong2025search}. To achieve this, we exert CA$_m$ and CA$_c$ to measure the correctness of translation results from both method-level and class-level with our manually crafted test suites. Besides, we also examine their CSR from a compilableness perspective. Moreover, we adopt DEP$_F$, DEP$_M$, and DEP$_L$ to investigate the ability of LLMs to sense contextual dependencies when translating code. We do not include literal consistency, e.g., exact match accuracy, for assessment because class-level code translation is way more challenging, and no sample can be translated exactly the same with ground truths, making such evaluation meaningless. In addition, semantically equivalent programs are not necessarily consistent in literal expressions. Hence, considering the above diverse evaluation workarounds, this threat is limited.   

\vspace{-1em}
\section{Conclusion and Future Work}
This work contributed to the first class-level code translation benchmark, namely ClassEval-T, and conducted extensive experiments with 6 recent LLMs of diverse categories and sizes. Results demonstrate that LLMs perform much worse on class-level code translations while neck-to-neck on a previous method-level benchmark, showing the motivation and necessity of constructing ClassEval-T. Afterwards, we further analyzed diverse translation strategies and their application scenarios, offering valuable guidance for practical usage. Finally, we proceeded with a thorough manual analysis and categorization of LLMs' translation failures on class-level code samples, as the first attempt, shedding light for researchers and practitioners to facilitate their studies and usage. 

{
Based on our research findings, our future work aims to enhance the performance of LLMs in class-level code translation tasks by integrating structural information (e.g., AST), to minimize syntax errors and improve translation accuracy. Additionally, we will extend ClassEval-T to include a broader range of PL pairs, those with strong industrial relevance, such as C++ to Rust.}

\noindent\textbf{Data Availability:} We open source the replication package and ClassEval-T benchmark at \cite{wLinHooC78:online}.

\vspace{-1em}
\section*{Acknowledgments}

This work was partially supported by the National Natural Science Foundation of China (Grant Nos. U24B20149  and 62272271), the Natural Science Foundation of Shandong Province (Grant No. ZR2024QF093), the Taishan Scholars Program (Grant No. tsqn202408009), the Research Grants Council of Hong Kong, and the Industry Research Project funds (Grant Nos. 6000871, 6000796, 9229109, 9229098, 9220103, and 9229029).

\bibliographystyle{ACM-Reference-Format}
\bibliography{sample-base}


\end{document}